\documentstyle[psfig]{mn}

\newif\ifAMStwofonts

\ifoldfss
  \ifCUPmtlplainloaded \else
    \NewTextAlphabet{textbfit} {cmbxti10} {}
    \NewTextAlphabet{textbfss} {cmssbx10} {}
    \NewMathAlphabet{mathbfit} {cmbxti10} {} 
    \NewMathAlphabet{mathbfss} {cmssbx10} {} 
  \fi
  \ifAMStwofonts
    \ifCUPmtlplainloaded \else
      \NewSymbolFont{upmath} {eurm10}
      \NewSymbolFont{AMSa} {msam10}
      \NewMathSymbol{\upi}     {0}{upmath}{19}
      \NewMathSymbol{\umu}     {0}{upmath}{16}
      \NewMathSymbol{\upartial}{0}{upmath}{40}
      \NewMathSymbol{\leqslant}{3}{AMSa}{36}
      \NewMathSymbol{\geqslant}{3}{AMSa}{3E}

      \let\leq=\leqslant \let\le=\leqslant
      \let\geq=\geqslant \let\ge=\geqslant
    \fi
  \fi
\fi 

\ifnfssone
  \newmathalphabet{\mathit}
  \addtoversion{normal}{\mathit}{cmr}{m}{it}
  \addtoversion{bold}{\mathit}{cmr}{bx}{it}
  \newmathalphabet{\mathbfit} 
  \addtoversion{normal}{\mathbfit}{cmr}{bx}{it}
  \addtoversion{bold}{\mathbfit}{cmr}{bx}{it}
  \newmathalphabet{\mathbfss} 
  \addtoversion{normal}{\mathbfss}{cmss}{bx}{n}
  \addtoversion{bold}{\mathbfss}{cmss}{bx}{n}
  \ifAMStwofonts
    \ifCUPmtlplainloaded \else
      \UseAMStwoboldmath
      \makeatletter
      \new@mathgroup\upmath@group
      \define@mathgroup\mv@normal\upmath@group{eur}{m}{n}
      \define@mathgroup\mv@bold\upmath@group{eur}{b}{n}
      \edef\UPM{\hexnumber\upmath@group}
      \new@mathgroup\amsa@group
      \define@mathgroup\mv@normal\amsa@group{msa}{m}{n}
      \define@mathgroup\mv@bold\amsa@group{msa}{m}{n}
      \edef\AMSa{\hexnumber\amsa@group}
      \makeatother
      \mathchardef\upi="0\UPM19
      \mathchardef\umu="0\UPM16
      \mathchardef\upartial="0\UPM40
      \mathchardef\leqslant="3\AMSa36
      \mathchardef\geqslant="3\AMSa3E

      \let\leq=\leqslant \let\le=\leqslant
      \let\geq=\geqslant \let\ge=\geqslant
    \fi
  \fi
\fi 

\ifnfsstwo
  \DeclareMathAlphabet{\mathbfit}{OT1}{cmr}{bx}{it}
  \SetMathAlphabet\mathbfit{bold}{OT1}{cmr}{bx}{it}
  \DeclareMathAlphabet{\mathbfss}{OT1}{cmss}{bx}{n}
  \SetMathAlphabet\mathbfss{bold}{OT1}{cmss}{bx}{n}
  \ifAMStwofonts
    \ifCUPmtlplainloaded \else
      \DeclareSymbolFont{UPM}{U}{eur}{m}{n}
      \SetSymbolFont{UPM}{bold}{U}{eur}{b}{n}
      \DeclareSymbolFont{AMSa}{U}{msa}{m}{n}
      \DeclareMathSymbol{\upi}{0}{UPM}{"19}
      \DeclareMathSymbol{\umu}{0}{UPM}{"16}
      \DeclareMathSymbol{\upartial}{0}{UPM}{"40}
      \DeclareMathSymbol{\leqslant}{3}{AMSa}{"36}
      \DeclareMathSymbol{\geqslant}{3}{AMSa}{"3E}

      \let\leq=\leqslant \let\le=\leqslant
      \let\geq=\geqslant \let\ge=\geqslant
    \fi
  \fi
\fi 

\ifCUPmtlplainloaded \else
  \ifAMStwofonts \else 
    \def\upi{\pi}
    \def\umu{\mu}
    \def\upartial{\partial}
  \fi
\fi

\title[Chemical evolution and joint formation of quasars and spheroids]
      {Chemical evolution in a model for the joint formation of quasars and 
       spheroids}
\author[D. Romano et al.]
       {Donatella Romano,$^1$\thanks{E-mail: \textsf{romano@sissa.it}} 
	Laura Silva,$^2$
	Francesca Matteucci$^{3, 1}$
	and Luigi Danese$^1$\\
        $^1$International School for Advanced Studies, SISSA/ISAS, 
	    Via Beirut 2-4, I-34014 Trieste, Italy\\ 
	$^2$Osservatorio Astronomico di Trieste,
            Via G.B. Tiepolo 11, I-34131 Trieste, Italy\\
	$^3$Dipartimento di Astronomia, Universit\`a di Trieste,
            Via G.B. Tiepolo 11, I-34131 Trieste, Italy}
\date{Accepted .
      Received ;
      in original form }

\pagerange{\pageref{firstpage}--\pageref{lastpage}}
\pubyear{2002}

\begin{document}

\maketitle

\label{firstpage}

\begin{abstract}
Direct and indirect pieces of observational evidence point to a strong 
connection between high-redshift quasars and their host galaxies. In the 
framework of a model where the shining of the quasar is the episode that stops 
the formation of the galactic spheroid inside a virialized halo, it has been 
proven possible to explain the submillimetre source counts together with their 
related statistics and the local luminosity function of spheroidal galaxies. 
The time delay between the virialization and the quasar manifestation required 
to fit the counts is short and incresing with decresing the host galaxy mass. 
In this paper we compute the detailed chemical evolution of gas and stars 
inside virialized haloes in the framework of the same model, taking into 
account the combined effects of cooling and stellar feedback. Under the 
assumption of negligible angular momentum, we are able to reproduce the main 
observed chemical properties of local ellipticals. In particular, by using the 
same duration of the bursts which are required in order to fit the 
submillimetre source counts, we recover the observed increase of the Mg/Fe 
ratio with galactic mass. Since for the most massive objects the assumed 
duration of the burst is $T_{burst}$ $\la$ 0.6 Gyr, we end up with a picture 
for elliptical galaxy formation in which massive spheroids complete their 
assembly at early times, thus resembling a monolithic collapse, whereas 
smaller galaxies are allowed for a more prolonged star formation, thus 
allowing for a more complicated evolutionary history. In the framework of the 
adopted scenario, only quasar activity can provide energies large enough to 
stop the star formation very soon after virialization in the most massive 
galactic haloes. The chemical abundance of the gas that we estimate at the end 
of the burst matches well the metallicity inferred from the quasar spectra. 
Therefore, the assumption that quasar activity interrupts the main episode of 
star formation in elliptical galaxies turns out to be quite reasonable. In 
this scenario, we also point out that non-dusty extremely red objects are the 
best targets for searching for high-redshift Type Ia supernovae.
\end{abstract}

\begin{keywords}
galaxies: elliptical and lenticular, cD -- galaxies: evolution -- galaxies: 
formation -- galaxies: ISM -- galaxies: stellar content -- nuclear reactions, 
nucleosynthesis, abundances
\end{keywords}

\section{Introduction}

A traditional view for the formation history of an elliptical galaxy stems 
from the Milky Way collapse model of Eggen, Lynden-Bell \& Sandage (1962): 
monolithic collapse and rapid star formation lead to a subsequent track known 
as `passive evolution' (i.e., without further star formation). This behaviour 
can be obtained if a `galactic wind' develops as a result of the energy 
injection into the gas by supernovae (SNe): the gas still present in the 
galaxy is swept away and the subsequent evolution is determined only by the 
amount of matter and energy which is restored into the interstellar medium 
(ISM) by the dying stars (Mathews \& Baker 1971; Larson 1974a; Ikeuchi 1977). 
The characteristic time-scales for the occurrence of the galactic winds depend 
on the previous star formation history, on the assumptions on the dark matter 
amount and distribution, and on the SN energetics. This scenario is known as 
{\it early monolithic collapse} (e.g., Larson 1974b; Larson \& Tinsley 1974; 
Arimoto \& Yoshii 1987; Matteucci \& Tornamb\`e 1987; Matteucci 1992, 1994; 
Bressan, Chiosi \& Fagotto 1994).

On the other hand, hierarchical clustering in the Universe is a fundamental 
process. The merging events, taking place over a major fraction of the 
cosmological time, have been associated with bursts of star formation in 
building-up galaxies in semi-analytical models (e.g., Kauffmann, Guiderdoni \& 
White 1994; Baugh, Cole \& Frenk 1996; Kauffmann 1996; Kauffmann \& Charlot 
1998).

A long-lasting star formation has been thoroughly ruled out, at least in giant 
ellipticals, on the basis of the chemical properties of the stellar population 
and the gas leftover by the process of star formation, which constrain the 
chemical enrichment history of these spheroids, and hence the mechanisms 
responsible for their formation (Bower, Lucey \& Ellis 1992; Ellis et al. 
1997; Bernardi et al. 1998; Thomas 1999; Thomas \& Kauffmann 1999). Studies on 
luminosity functions at substantial redshift show that the majority of 
luminous E and S0 galaxies were already in place at $z$ $\simeq$ 1 and that 
most of their stars were relatively old (Im et al. 2001; Cohen 2001). Searches 
for a population of faint intrinsically red objects, representing the expected 
$z$ $>$ 1 precursors of passively evolving ellipticals which formed their 
stars at high redshift, have been conducted: Daddi et al. (2000) find that 
colours and surface density of extremely red objects (EROs) are consistent 
with the hypothesis that they are the precursors of the present day luminous 
elliptical galaxies. Also the clustering properties support this conclusion 
(Daddi et al. 2000; McCarthy et al. 2000; Magliocchetti et al. 2001). Several 
authors (e.g., Franceschini et al. 1991, 1994; Lilly et al. 1999; Dunlop 2001; 
Granato et al. 2001) suggested that the burst, during which most of the stars 
in spheroidal galaxies were formed at $z$ $>$ 1, is elusive in the optical 
bands, but shows up in the far-IR domain. Evidence of this phase has been 
found by Smail, Ivison \& Blain (1997) as a result of a deep survey with {\it 
SCUBA} and has been confirmed by subsequent surveys and follows-up (Hughes et 
al. 1998; Barger et al. 1998; Blain et al. 1999; Smail et al. 2000; Almaini et 
al. 2001; Borys et al. 2001).

The above-sketched picture is complemented by additional pieces of 
observational evidence pointing to a strong connection between quasi-stellar 
objects (QSOs) and galaxies. In particular, the recently discovered 
correlation between the central massive dark object (MDO) (routinely 
interpreted as a dormant black hole) and the hot stellar component of nearby 
galaxies (e.g., Kormendy \& Richstone 1995; Magorrian et al. 1998; van der 
Marel 1999; Ferrarese \& Merritt 2000; Gebhardt et al. 2000; McLure \& Dunlop 
2001; Merritt \& Ferrarese 2001a,b) suggests a direct connection between QSO 
activity and galaxy formation. Since $\sim$ 97 per cent of early-type galaxies 
do have MDOs (Magorrian et al. 1998), it seems reasonable to construct models 
taking into account jointly the cosmological formation of QSOs and spheroids 
(Silk \& Rees 1998; Fria\c ca \& Terlevich 1998; Kauffmann \& Haehnelt 2000; 
Monaco, Salucci \& Danese 2000; Granato et al. 2001, among others). While QSO 
activity and evolution at low and intermediate redshifts are thought to be 
driven by interactions between galaxies (Barnes \& Hernquist 1991), the QSO 
evolution at $z$ $>>$ 1 is expected to be more closely related to the 
formation of massive galaxies in overdense regions (Haehnelt \& Rees 1993). In 
particular, Granato et al. (2001) showed that most of the above-mentioned 
observations can be explained in a hierarchical model for spheroidal galaxy 
formation, if stellar and QSO feedback are properly included. In particular, 
they pointed out that the stellar feedback is more effective in reducing the 
star formation rate of smaller haloes, thus producing a star formation rate 
which is increasing with the halo mass and the final mass in stars.

The present study aims to investigate further the evolutionary histories for 
elliptical galaxies within the framework suggested by Granato et al. (2001), 
focusing on the expected chemo-photometric properties of the resulting stellar 
populations and on the chemistry of the gas. The predictions of the model are 
then tested against the body of the available data. We assume that after a 
vigorous star formation building up the bulk of the stellar population and 
enriching the gaseous medium to roughly solar or higher metallicity, the QSO 
shines at the centre, ionizing the surrounding medium and inhibiting further 
star formation. A galactic wind is possibly triggered at this point, due to 
the combined feedback of QSO and stars. The QSOs shine in an {\it inverted 
hierarchical order} (Monaco et al. 2000), which means that the time interval 
between the onset of the star formation and the peak of the QSO activity is 
shorter for more massive spheroids.

The paper is organized as follows: in \S 2 we report on the current status 
of observations, as far as both the integrated properties of the composite 
stellar population and the interstellar/surrounding gaseous component are 
concerned. In \S 3 we introduce our model and in \S 4 we present and compare 
the model results to the available observations. A summary is given in \S 5, 
together with a comparison with previous studies. In the following we will 
assume $\Omega_\Lambda$ = 0.7, $\Omega_M$ = 0.3 and $h$ = 0.7, unless 
otherwise stated.

\section{Observational constraints}

\subsection{The stellar populations}

Most of the information on elliptical galaxies comes from their integrated 
properties: abundances are derived either through colours or integrated 
spectra. The metallicity indices Mg$_2$ and $\langle$Fe$\rangle$, as 
originally defined in Faber, Burstein \& Dressler (1977) and Faber et al. 
(1985), are the most widely used metallicity indicators and have been measured 
for a number of objects. The integrated properties of elliptical galaxies can 
be analysed by means of population synthesis techniques (e.g., Buzzoni, 
Gariboldi \& Mantegazza 1992; Bruzual \& Charlot 1993; Bressan, Chiosi \& 
Fagotto 1994; Bressan, Chiosi \& Tantalo 1996; Tantalo, Chiosi \& Bressan 
1998) in order to get an estimate of the real abundances. Positive values of 
the elemental ratio [Mg/Fe] have been found in the nuclei of ellipticals, 
increasing with increasing the galactic mass (O'Connell 1976; Peterson 1976; 
Faber, Worthey \& Gonz\'alez 1992; Worthey, Faber \& Gonz\'alez 1992; Weiss, 
Peletier \& Matteucci 1995; Tantalo et al. 1998; Worthey 1998; J\o rgensen 
1999; Kuntschner 2000; Kuntschner et al. 2001; Terlevich \& Forbes 2001). 
Although large uncertainties are involved with these methods, the 
overabundance of the [Mg/Fe] ratio in ellipticals with respect to the solar 
value is accepted and generally interpreted as due to a short, intense star 
formation, perhaps coupled to an initial mass function (IMF) biased towards 
massive stars. Other observational clues consistent with this picture are 
reminded below:

Bender, Burstein \& Faber (1992) studied the distribution of elliptical 
galaxies in the Virgo cluster and in the Coma cluster in a 3-space of the 
basic global parameters central velocity dispersion $\sigma$, effective 
radius $r_e$, and mean effective surface brightness $I_e$. They concluded that 
cluster ellipticals lie on a fundamental plane (FP, Djorgovski \& Davis 1987), 
i.e., they do not fill uniformly the space of the global parameters but rather 
select a bidimensional region of it (see also Dressler et al. 1987). The 
tightness of the FP for elliptical galaxies in local clusters was then 
confirmed by other works (e.g., Renzini \& Ciotti 1993). It points to a high 
uniformity and synchronization in the galaxy formation process in rich 
clusters.

Bower et al. (1992) pointed out the existence of a very tight 
colour\,--\,$\sigma$ relation for elliptical galaxies in the Virgo and Coma 
clusters. By combining these cluster data at low redshift with 
{\it HST}-selected samples at intermediate redshift, it was demonstrated that 
{\it at least cluster ellipticals} are made of very old stars, with the bulk 
of them having formed at redshift $z$ $\ge$ 2 (Ellis et al. 1997). A tight 
colour\,--\,magnitude relation (CMR) was also found, in the sense that colours 
become redder with increasing luminosity. The tightness of the CMR for 
elliptical galaxies in clusters has been confirmed up to $z$ $\sim$ 1 
(Aragon-Salamanca et al. 1993; Stanford, Eisenhardt \& Dickinson 1998). The 
optical-IR colours of the early-type cluster galaxies observed by Stanford et 
al. (1998) become bluer with increasing $z$ (out to $z$ = 0.9) in a manner 
consistent with the passive evolution of an old stellar population formed at 
an early cosmic epoch. Moreover, the modest shift with increasing $z$ found in 
the zero-point of the FP, Mg$_2$\,--\,$\sigma$, and CMRs of cluster 
ellipticals (e.g., Dickinson 1995; Ellis et al. 1997; Ziegler \& Bender 1997; 
Bender et al. 1998; van Dokkum et al. 1998; Kodama et al. 1998; Stanford et 
al. 1998) has leaded to the conclusion that most stars in cluster ellipticals 
formed at $z$ $\ge$ 3. 

However, until a couple of years ago much of this evidence was restricted to 
cluster elliptical galaxies. Using new observations for a sample of 931 
early-type galaxies assigned to three different environments (clusters, 
groups, and field), Bernardi et al. (1998) demonstrated that cluster, group, 
and field early-type galaxies follow almost identical Mg$_2$\,--\,$\sigma$ 
relations. The largest Mg$_2$ zero-point difference they found is only 0.007 
$\pm$ 0.002 mag, thus implying an age difference of only $\sim$ 1 Gyr between 
the stellar populations of cluster and field elliptical galaxies, these latter 
being younger on the average. Recently, it has been claimed that significant 
age variations might be hidden in the Mg$_2$\,--\,$\sigma$ relation, so that 
it would be desirable to consider the two Balmer line indices H$\beta$ and 
Hn/Fe as more robust age indicators (Concannon, Rose \& Caldwell 2000). 
However, the scatter of H$\beta$ observed in cluster and luminous field 
elliptical galaxies can also be explained by composite populations that 
contain a small fraction of old metal-poor stars, without requiring a young 
stellar component (Maraston \& Thomas 2000). On the other hand, there is 
evidence in both cluster and field ellipticals that a modest fraction of the 
present day stellar population is indeed formed in secondary bursts at 
relatively low redshift (e.g., Trager et al. 2000a,b; Treu et al. 2001).

Another piece of observational evidence supporting a scenario where an early, 
intense star formation is followed by a longer, quiescent phase (passive 
evolution) is given by the present day Type Ia and II SN rates in early-type 
galaxies: $Rate_{SNIa}$ = 0.18 $\pm$ 0.06 SNu\footnote{\small{1 SNu = 1 SN per 
century per 10$^{10}$ $L_{B\,\odot}$.}}, $Rate_{SNII}$ $<$ 0.02 SNu (for 
$H_\circ$ = 75 km s$^{-1}$ Mpc$^{-1}$; Cappellaro, Evans \& Turatto 1999). 
Since Type II SNe come from short-living progenitors, whereas Type Ia SNe come 
from long-living ones, the observed Type Ia and II SN rates imply that the 
star formation in early-type galaxies must be inactive at the present time.

\subsection{The gaseous medium}

\subsubsection{The quasar environment}

In a recent paper, Hamann \& Ferland (1999) have reviewed the topic of QSO 
emission and intrinsic absorption lines. They have shown that these lines can 
give us some hints on the chemical composition of the QSO environment. The 
metallicity of broad emission line regions (BELRs) can be studied by using 
emission ratios. Hamann et al. (2001) find evidence for roughly solar or 
higher metallicities out to $z$ $>$ 4 and enhanced N/C ratios with respect to 
solar in more luminous objects (confirmed by Fan et al.'s 2001 spectral 
analysis for a few $z$ $\sim$ 6 quasars). This implies that most of the 
enrichment of the gaseous medium must occur before the QSOs become observable 
(i.e., on time-scales $\le$ 1 Gyr, at least at the highest redshifts). This 
gives further support to the idea that QSOs are hosted at the centres of 
elliptical galaxies that experienced vigorous star formation at some early 
stage of their evolution.

\subsubsection{The hot gas}

The standard SN rates predict the iron abundance of the ISM in early-type 
galaxies to be as high as several times the solar value (e.g., 
Loewenstein \& Mathews 1991; Ciotti et al. 1991; Renzini et al. 1993). 
However, the first measurements of the ISM with {\it ASCA} showed that 
the metallicity was less than half a solar (Awaki et al. 1994; Loewenstein 
et al. 1994; Mushotzky et al. 1994; Matsushita et al. 1994; Arimoto et 
al. 1997; Matsumoto et al. 1997). Giant early-type galaxies are estimated 
to have a roughly solar stellar iron abundance (e.g., Arimoto et al. 1997). 
Thus, the X-ray measured abundances implied that the ISM metal abundance was 
even lower than the stellar metallicity. Lately, from {\it ASCA} and 
{\it ROSAT} spectra of four among the brightest elliptical galaxies in X-rays, 
Fe abundances of $\sim$ 1\,--\,2 times solar and (except for one object) 
relative abundances fixed at their solar values have been found (Buote 
1999)\footnote{\small{The Fe abundances obtained by Buote should be increased 
by a factor of 1.44 to reflect the meteoritic solar abundances instead of the 
photospheric values he used (see Ishimaru \& Arimoto 1997).}}. This result has 
been confirmed by the analysis of {\it ASCA} data for 27 giant early-type 
galaxies by Matsushita, Ohashi \& Makishima (2000): they find nearly solar 
(within a factor of 2) abundances of Fe and $\alpha$-elements in X-ray 
luminous galaxies. Much more room for uncertainty is left for X-ray fainter 
galaxies, but it seems that the contribution from Type Ia SNe to the ISM 
abundance is lower in X-ray faint than in X-ray luminous systems. These 
results strongly suggest that a large fraction of the SNIa products have 
escaped into the intergalactic space.

An additional clue to the possible relation between QSOs and galaxy formation 
comes from X-ray observations of the intergalactic medium (IGM) in galaxy 
clusters. As soon as deviations from the expected $L_X \propto T_v^2$ relation 
emerged from the cluster data, several authors suggested a possible 
pre-heating of the intergalactic gas (e.g. Kaiser 1991; Evrard \& Henry 1991). 
Evidence of an entropy excess with respect to the values expected from purely 
gravitational heating has been recently found by Ponman, Cannon \& Navarro 
(1999) and Lloyd-Davies, Ponman \& Cannon (2000), although the interpretation 
of the data is still debated (see, e.g., Roussel, Sadat \& Blanchard 2000). 
The extra-energy injected is in the range $\sim$ 0.5\,--\,3 keV per particle 
(e.g., Cavaliere, Menci \& Tozzi 1997; Wu, Fabian \& Nulsen 2000; Borgani et 
al. 2001). On one hand, the most natural heating mechanism, heating from SNe, 
seems to fall short of the required energy budget (Valageas \& Silk 1999; 
Kravtsov \& Yepes 2000 and refs. therein; but see also Voit \& Bryan 2001). On 
the other hand, QSOs/AGNs could supply at least part of such a large amount of 
energy per particle, provided that a fraction of their total energy output is 
transferred to the IGM by some physical mechanism, such as, for instance, 
heating by outflows and jets, as suggested by several authors (e.g., Valageas 
\& Silk 1999; Inoue \& Sasaki 2001; Yamada \& Fujita 2001). Interactions 
between radio lobes of powerful radiosources and the X-ray emitting IGM in 
clusters have been recently studied with Chandra (Fabian et al. 2000; Fabian 
et al. 2001).

\section{The model}

\subsection{Basic assumptions}

Since we are mostly interested in understanding spheroidal galaxies with 
substantial mass ($M_{sph}$ $\ga$ 10$^{10}$ $M_\odot$) formed at high 
redshift, we concentrate on star formation inside large virialized haloes of 
dark matter (DM). At the beginning, baryonic and DM share the same density 
profile. Then, baryonic matter cools, collapses and starts forming stars. A 
single zone ISM with instantaneous mixing of gas is assumed through the inner 
region where the processes of star formation from cold gaseous clouds, 
feedback from SNe, and accretion of mass by infall from the surrounding, 
cooling medium are taking place. The star formation is a very efficient 
process, building up the bulk of the stellar population on a very short 
time-scale, until the QSO shines at the centre. At that time the star 
formation stops and the galaxy undergoes a subsequent phase of passive 
evolution -- Type Ia SNe, exploding in an already ionized medium since then, 
are very efficient in keeping hot the gas (Recchi, Matteucci \& D'Ercole 
2001). The age of the galaxy models, $T_{gal}$, is a function of the adopted 
cosmology and redshift of galaxy formation, $z_f$.

Following Navarro, Frenk \& White (1997) and the generalization by Bullock et 
al. (2001), a virialized dark halo of mass $M_{vir}$ identified at $z$ = $z_0$ 
has a virial radius defined by
\begin{equation}
M_{vir}=\frac{4 \pi}{3} \Delta_{vir}(z_0)\rho_u(z_0) r_{vir}^3,
\end{equation}
where $\rho_u(z_0)$ is the mean universal density and $\Delta_{vir}(z_0)$ is 
the virial overdensity at that redshift. For flat cosmologies a good 
approximation has been derived by Bryan \& Norman (1998):
\begin{equation}
\Delta_{vir}\simeq \frac{18 \pi^2 + 82x-39x^2}{\Omega(z)},
\end{equation}
where $x$ = $\Omega(z)-1$ and $\Omega(z)$ is the ratio of the mean matter 
density to the critical density at redshift $z$.

We can also define a virial velocity, $V^2_{vir}$ = $G M_{vir}/r_{vir}$, and 
the temperature of the gas in hydrostatic equilibrium,
\begin{equation}
kT=\frac{1}{2} \mu m_p V_{vir}^2,
\end{equation}
where $\mu m_p$ is the mean molecular weight of the gas.

Following numerical experiments (see, e.g., Navarro, Frenk \& White 1996, 
1997; Bullock et al. 2001), the density profile within a virialized halo is 
well represented by:
\begin{equation}
\rho(r) = \frac{\rho_s}{c\,x(1 + c\,x)^2},
\end{equation}
where $x$ = $r/r_{vir}$ and $c$ is the concentration parameter. Integrating 
the density profile over the radius up to the virial radius we get the halo 
mass:
\begin{equation}
M_{vir}=4 \pi \rho_s r_{vir}^3 c^{-3} \left[\ln(1+c)-\frac{c}{1+c} \right].
\end{equation}
Numerical simulations (Navarro et al. 1997; Bullock et al. 2001) show that the 
concentration parameter $c$ is a  function of the mass ($c$ $\propto$ 
$M_{vir}^{-0.1}$) and of the redshift [$c$ $\propto$ (1 + $z_0$)$^{-1}$]. 
These dependences have been included in the model.

In this model we assume that the baryon component in virialized haloes has 
initially the same distribution as the DM. However, baryons cool down in a 
characteristic time defined at each radius through the ratio of specific 
energy content to cooling rate,
\begin{equation}
t_{cool}(r) = \frac{3}{2} \frac{\rho_{gas}(r)}{\mu m_{p}} 
\frac{k T}{n^2_{e}(r) \Lambda(T)},
\end{equation}
where $\rho_{gas}(r)$ is the gas density, $n_{e}(r)$ is the electron density, 
$m_{p}$ is the proton mass, $\mu m_{\mathrm{p}}$ is the mean molecular weight 
of the gas and $\Lambda(T)$ is a suitable cooling function. In the following 
we adopt cooling functions taken from Sutherland \& Dopita (1993), which 
include the dependence on metal abundance.

The second relevant time is the dynamical time, $t_{dyn}(r)$ = $[ 3 \pi /32 G 
\rho(r) ]^{1/2}$. We neglect the angular momentum of the baryons. Actually, 
the linear tidal torque theory shows that later collapsing shells had more 
time to gain spin via tidal torques. As a result, the distribution of the spin 
parameter exhibits a log normal distribution with the mean value of the spin 
parameter $\lambda$ and its standard deviation $\sigma_{\lambda}$ 
significantly decreasing with increasing redshift (see, e.g., Maller, Dekel \& 
Somerville 2002). Therefore, a large fraction of low-spin objects are expected 
among the haloes virialized at high redshift, $z$ $\ga$ 2\,--\,3, which in our 
model are the hosts of the present day massive spheroids (see below).

The third time, $T_{burst}$ = $t_{QSO}-t_{vir}$, is the time delay between the 
virialization time $t_{vir}$, which we assume to be also the time $t_f$ when 
stars begin to form in the galaxy, and the time of the shining of the QSO 
$t_{QSO}$, which sets the end of the formation of the bulk of the stars. The 
time $T_{burst}$ is directly related to the visibility of the 
proto-ellipticals in the far-IR. As shown by Granato et al. (2001), the 850 
$\mu$m source counts down to 0.5 mJy and related statistics are well 
reproduced by assuming that: 

\begin{equation}
T_{burst}(M_{sph})=\left\{ \begin{array}{ll}
T_{b}^{*}, & {\mathrm if}\, M_{sph}\geq M^{*}_{sph}\\
T_{b}^{*}+ \log \frac{M^{*}_{sph}}{M_{sph}}, & {\mathrm if}\,
M_{sph}\leq M^{*}_{sph}
\end{array}\right.
\label{eq:delay1} 
\end{equation}

\noindent where $T_{b}^*$ $\sim$ 0.5 Gyr is the burst time-scale for galaxies 
with $M_{sph}$ $\geq$ $M_{sph}^*$, and $M_{sph}^*$ $\sim$ 1.5 $\times$ 
10$^{11}$ $M_{\odot}$ is the characteristic mass in the local stellar mass 
function of galaxies.

For each halo of mass $M_{vir}$ virialized at redshift $z_{vir}$ we can define 
a radius $r_{coll}$, as the radius within which 
\begin{equation}
{\mathrm max}[t_{cool}(r),t_{dyn}(r)] \leq T_{burst}.  
\end{equation}
The radius $r_{coll}$ coincides with $r_{vir}$ for small mass haloes, while it 
is significantly smaller for larger haloes. This is due to the increase of the 
cooling time with increasing halo mass and produces a natural cut-off in the 
stellar mass in galaxies.

We can also define the time-scale of the gas infall, $\tau_{inf}$ = 
max[$t_{cool}(r_{coll})$, $t_{dyn}(r_{coll})$].

The fundamental equations of chemical evolution can be written as:
\begin{displaymath}
\frac{{\mathrm{d}}G_i(t)}{{\mathrm{d}}t} = -X_i(t)\psi(t) + R_i(t) +
\Bigg( \frac{{\mathrm{d}}G_i}{{\mathrm{d}}t} \Bigg)_{inf}
\end{displaymath}
\begin{equation}
\hspace{1.2cm} - \Bigg( \frac{{\mathrm{d}}G_i}{{\mathrm{d}}t} \Bigg)_{reh}
\end{equation}
where
\begin{displaymath}
R_i(t) = \int_{M_{low}}^{M_{B_{min}}} \psi(t - \tau_{M})Q_{Mi}(t - 
\tau_{M})\phi(M) {\mathrm{d}}M +
\end{displaymath}
\begin{displaymath}
\hspace{1cm} A \int_{M_{B_{min}}}^{M_{B_{max}}}\phi(M_B)
\Bigg\{ \int_{\mu_{min}}^{0.5} f(\mu) \psi(t - \tau_{M_2})
\end{displaymath}
\begin{displaymath}
\hspace{1cm} Q_{M_1i}(t - \tau_{M_2}) {\mathrm{d}}\mu \Bigg\} {\mathrm{d}}M_B 
+
\end{displaymath}
\begin{displaymath}
\hspace{1cm} (1 - A) \int_{M_{B_{min}}}^{M_{B_{max}}} \psi(t - 
\tau_{M})Q_{Mi}(t - \tau_{M})\phi(M) {\mathrm{d}}M +
\end{displaymath}
\begin{equation}
\hspace{1cm} \int_{M_{B_{max}}}^{M_{up}} \psi(t - 
\tau_{M})Q_{Mi}(t - \tau_{M})\phi(M) {\mathrm{d}}M.
\end{equation}
$G_i(t)$ = $X_i(t)\,M_{cold}(t)$ is the cold gas mass in form of the element 
$i$. The quantity $X_i(t)$ represents the abundance by mass of the element $i$ 
(by definition, the summation over all the elements present in the gas mixture 
is equal to unity). The various integrals of Eq.(10) represent the rates at 
which SNe (I and II) as well as single low- and intermediate-mass stars 
restore their processed and unprocessed material to the ISM (see Matteucci \& 
Greggio 1986 for details).

The star formation rate (SFR) is given by:
\begin{equation}
\psi(t) = \int_0 ^{r_{coll}} \frac{1}{{\mathrm max}[t_{cool}(r),t_{dyn}(r)]} 
\frac {{\mathrm d}M^k_{cold}(r, t)}{{\mathrm d}r} {\mathrm d}r,
\end{equation}
where $M_{cold}(t)$ is the gas mass which is cold by time $t$ and $k$ = 1. We 
assume that the gas distribution follows the DM distribution during the burst 
of star formation. Of course, the above formula can be rearranged in the form 
$\psi(t)$ = $M^k_{cold}(t)/\tau_{\star}$ = $\nu_{\star} M^k_{cold}(t)$, with 
$\tau_{\star}$ being the appropriate time-scale for star formation and 
$\nu_{\star}$ the corresponding star formation efficiency.

The infall term is expressed as:
\begin{equation}
\Bigg( \frac{{\mathrm{d}}G_i}{{\mathrm{d}}t} \Bigg)_{inf} = 
(X_i)_{inf} \frac{M_{gas}}{\tau_{inf}} \frac{1}{1 - 
{\mathrm e}^{-T_{burst}/\tau_{inf}}} {\mathrm e}^{-t/\tau_{inf}}
\end{equation}
where $(X_i)_{inf}$ = $(X_i)_{\mathrm{P}}$, i.e., the infalling gas has a 
primordial chemical composition, and $M_{gas}$ is the gas mass within 
$r_{coll}$. In this formulation the accretion lasts during all the burst, 
before the QSO shines.

The last term of Eq.(9) is the rate of reheating, which accounts for feedback 
from SNe. It gives the amount of cold gas which is heated and subtracted to 
further stellar processing per unit time due to SN explosions:
\begin{equation}
\Bigg( \frac{{\mathrm{d}}G_i}{{\mathrm{d}}t} \Bigg)_{reh} = X_i(t) \psi(t) \, 
\epsilon \, \frac{4}{5} \, \frac{\eta_{SN} E_{SN}}{V_{vir}^2}
\end{equation}
(Kauffmann, White \& Guiderdoni 1993). $\eta_{SN}$ is the number of Type II 
SNe expected per solar mass of stars formed and is computed according to the 
particular choice of the IMF, $E_{SN}$ is the kinetic energy of the ejecta 
from each supernova ($\sim$ 10$^{51}$ erg) and $\epsilon$ is the fraction of 
this energy which is used to reheat the cold gas to the virial temperature of 
the halo. It is worth noticing that we neglect the contribution of Type Ia SNe 
to the reheating. In fact, during the burst the main contribution in terms of 
energy is coming from Type II SNe, and the end of the burst is set by the 
onset of the QSO activity. Thus, we expect that most of Type Ia SNe explode in 
a warm, rarefied medium. Therefore, they play their major role in polluting 
the intracluster medium (ICM) with their end products at later times, rather 
than efficiently regulate the process of star formation at former times.

\begin{table*}
\centering
\begin{minipage}{15.5cm}
\caption{Mg and Fe isotopes in the ejecta of core-collapse supernovae (Nomoto 
	 et al. 1997).}
\begin{tabular}{@{}cccccccc}
\hline
$M_{init}$ & $M^{ej}_{^{24}Mg}$ & $M^{ej}_{^{25}Mg}$ & $M^{ej}_{^{26}Mg}$ &
$M^{ej}_{^{54}Fe}$ & $M^{ej}_{^{56}Fe}$ & $M^{ej}_{^{57}Fe}$ & 
$M^{ej}_{^{58}Fe}$ \\
\hline
 13 & 9.23 $\times$ 10$^{-3}$ & 1.38 $\times$ 10$^{-3}$	& 8.96 $\times$ 
10$^{-4}$ & 2.10 $\times$ 10$^{-3}$ & 1.50 $\times$ 10$^{-1}$ & 4.86 $\times$ 
10$^{-3}$ & 3.93 $\times$ 10$^{-9}$ \\
 15 & 3.16 $\times$ 10$^{-2}$ & 2.55 $\times$ 10$^{-3}$	& 2.03 $\times$ 
10$^{-3}$ & 4.49 $\times$ 10$^{-3}$ & 1.44 $\times$ 10$^{-1}$ & 4.90 $\times$ 
10$^{-3}$ & 1.27 $\times$ 10$^{-8}$ \\
 18 & 3.62 $\times$ 10$^{-2}$ & 7.54 $\times$ 10$^{-3}$	& 5.94 $\times$ 
10$^{-3}$ & 6.04 $\times$ 10$^{-3}$ & 7.57 $\times$ 10$^{-2}$ & 2.17 $\times$ 
10$^{-3}$ & 1.37 $\times$ 10$^{-8}$ \\
 20 & 1.47 $\times$ 10$^{-1}$ & 1.85 $\times$ 10$^{-2}$	& 1.74 $\times$ 
10$^{-2}$ & 2.52 $\times$ 10$^{-3}$ & 7.32 $\times$ 10$^{-2}$ & 3.07 $\times$ 
10$^{-3}$ & 3.70 $\times$ 10$^{-9}$ \\
 25 & 1.59 $\times$ 10$^{-1}$ & 3.92 $\times$ 10$^{-2}$	& 3.17 $\times$ 
10$^{-2}$ & 4.81 $\times$ 10$^{-3}$ & 5.24 $\times$ 10$^{-2}$ & 1.16 $\times$ 
10$^{-3}$ & 8.34 $\times$ 10$^{-9}$ \\
 40 & 3.54 $\times$ 10$^{-1}$ & 4.81 $\times$ 10$^{-2}$	& 1.07 $\times$ 
10$^{-1}$ & 9.17 $\times$ 10$^{-3}$ & 7.50 $\times$ 10$^{-2}$ & 2.29 $\times$ 
10$^{-3}$ & 1.29 $\times$ 10$^{-8}$ \\
 70 & 7.87 $\times$ 10$^{-1}$ & 1.01 $\times$ 10$^{-1}$	& 2.91 $\times$ 
10$^{-1}$ & 5.81 $\times$ 10$^{-3}$ & 7.50 $\times$ 10$^{-2}$ & 3.83 $\times$ 
10$^{-3}$ & 4.17 $\times$ 10$^{-8}$ \\
100 & 1.22                    & 1.50 $\times$ 10$^{-1}$	& 4.80 $\times$ 
10$^{-1}$ & 5.81 $\times$ 10$^{-3}$ & 7.50 $\times$ 10$^{-2}$ & 3.83 $\times$ 
10$^{-3}$ & 4.17 $\times$ 10$^{-8}$  \\
\hline
\end{tabular}
\end{minipage}
\end{table*}

An important parameter entering all chemical evolution models is the IMF 
slope, which describes the relative numbers of stars born at a given mass. In 
spite of recent observational progress, many fundamental properties of the IMF 
are still unknown. In particular, there is no consensus on the question of 
whether the IMF is independent of environmental conditions such as stellar 
density and metallicity. New investigations have come to different conclusions 
on the universality of the IMF (Eisenhauer 2001; Gilmore 2001; Kroupa 2001); 
nevertheless it seems to exist some empirical indication that the IMF is 
systematically biased towards more massive stars in the early Universe and in 
starbursts (Eisenhauer 2001 and references therein). We will consider three 
cases: {\it i)} a Salpeter (1955) IMF, $\phi(M)$ $\propto$ $M^{-1.35}$, {\it 
ii)} an IMF slightly biased towards more massive stars, $\phi(M)$ $\propto$ 
$M^{-1.15}$, and {\it iii)} a two slope IMF, $\phi(M)$ $\propto$ $M^{-0.4}$ 
for $M$ $\le$ 1 $M_\odot$ and $\phi(M)$ $\propto$ $M^{-1.25}$ for $M$ $>$ 1 
$M_\odot$, all normalized to 1 over the mass range 0.1\,--\,100 $M_\odot$. 
Besides, we will assume that the IMF {\it is not} varying with time, since 
this seems to be the case for our own galaxy (Chiappini, Matteucci \& Padoan 
2000).

\begin{figure}
\centerline{\psfig{figure=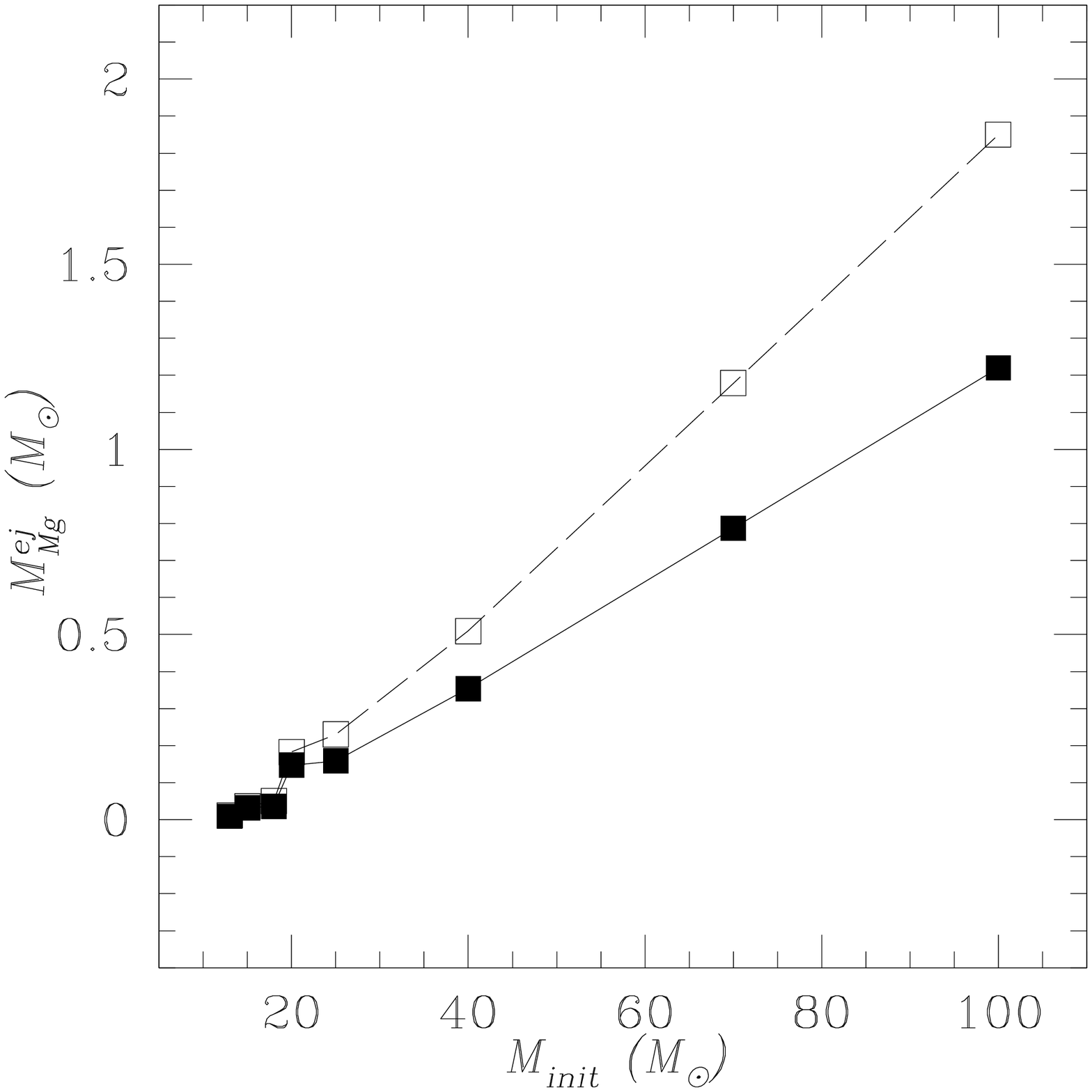,width=8.3cm,height=8.3cm} }
\caption{Masses ejected in form of $^{24}$Mg (solid squares and continuous 
line) and in form of $^{24}$Mg + $^{25}$Mg + $^{26}$Mg (open squares and 
dashed line) as a function of the initial mass of the progenitor star (the 
values relevant to the 100 $M_\odot$ star have been extrapolated). Yields from 
Nomoto et al. (1997).}
\end{figure}

\begin{figure}
\centerline{\psfig{figure=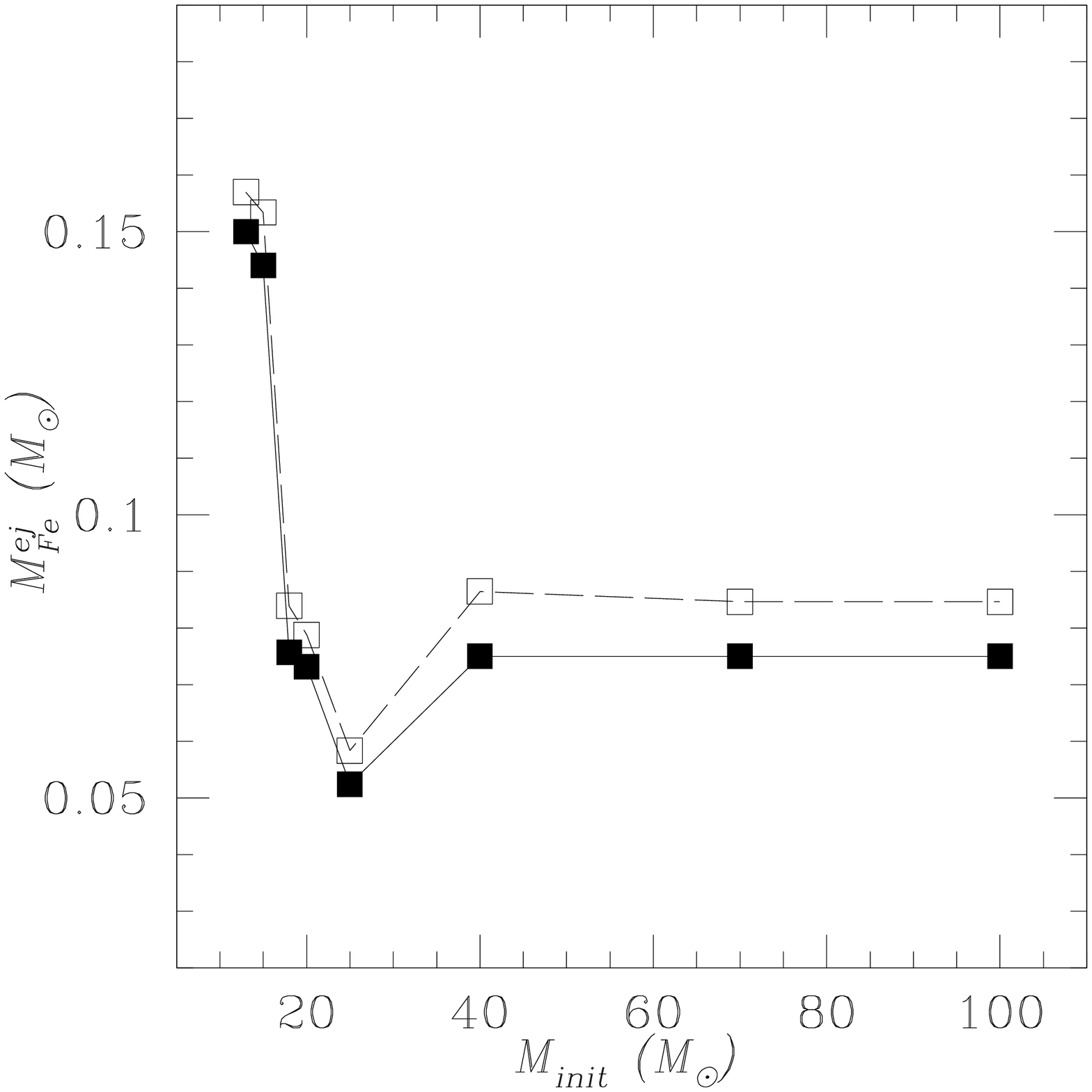,width=8.3cm,height=8.3cm} }
\caption{Masses ejected in form of $^{56}$Fe (solid squares and continuous 
line) and in form of $^{54}$Fe + $^{56}$Fe + $^{57}$Fe + $^{58}$Fe (open 
squares and dashed line) as a function of the initial mass of the progenitor 
star (the values relevant to the 100 $M_\odot$ star have been extrapolated). 
Yields from Nomoto et al. (1997).}
\end{figure}

\begin{table*}
\centering
\begin{minipage}{15.5cm}
\caption{Model parameters: dark halo mass, baryonic mass inside the halo, 
         fraction of the baryonic mass which cools and collapses, fraction of 
         the kinetic energy from Type II SNe which reheates the gaseous 
         medium, circular velocity of the halo at the time of galaxy 
         formation, gas temperature at the time of galaxy formation, 
         efficiency of star formation, duration of the burst of star 
         formation. The parameters listed here are relevant to the case $z_f$ 
         = 5. The masses are expressed in units of $M_\odot$, $V_{vir}$ is in 
         units of km s$^{-1}$, $T$ is in units of K, $\nu_\star$ is in units 
         of Gyr$^{-1}$, and $T_{burst} = t_{QSO}-t_f$ is in units of Gyr. See 
         text for more details.}
\begin{tabular}{@{}ccccccccc}
\hline
Model & $M$ & $M_{bar}$ & $\alpha$ & $\epsilon$ & $V_{vir}$ & $T$ & 
$\nu_\star$ & $T_{burst}$ \\
\hline
1{\it a} & 1.65 $\times$ 10$^{13}$ & 2.47 $\times$ 10$^{12}$ & 0.21 & 0.15 & 
        749.4 & 2.04 $\times$ 10$^{7}$ & 5.94 & 0.60 \\
2{\it a} & 2.00 $\times$ 10$^{12}$ & 3.00 $\times$ 10$^{11}$ & 0.77 & 0.15 & 
        370.9 & 4.99 $\times$ 10$^{6}$ & 7.27 & 0.86 \\
3{\it a} & 6.00 $\times$ 10$^{11}$ & 9.00 $\times$ 10$^{10}$ & 1.00 & 0.15 & 
        248.3 & 2.24 $\times$ 10$^{6}$ & 8.40 & 1.17 \\
4{\it a} & 3.00 $\times$ 10$^{11}$ & 4.50 $\times$ 10$^{10}$ & 1.00 & 0.15 & 
        197.1 & 1.41 $\times$ 10$^{6}$ & 8.63 & 1.56 \\
5{\it a} & 2.00 $\times$ 10$^{11}$ & 3.00 $\times$ 10$^{10}$ & 1.00 & 0.15 & 
        172.2 & 1.08 $\times$ 10$^{6}$ & 8.77 & 1.86 \\
6{\it a} & 1.26 $\times$ 10$^{11}$ & 1.89 $\times$ 10$^{10}$ & 1.00 & 0.15 & 
        147.6 & 7.91 $\times$ 10$^{5}$ & 8.93 & 2.09 \\
7{\it a} & 7.50 $\times$ 10$^{10}$ & 1.13 $\times$ 10$^{10}$ & 1.00 & 0.15 & 
        124.2 & 5.59 $\times$ 10$^{5}$ & 9.13 & 2.47 \\
8{\it a} & 2.10 $\times$ 10$^{10}$ & 3.15 $\times$ 10$^{9}$ & 1.00 & 0.15 & 
        81.2 & 2.39 $\times$ 10$^{5}$ & 9.65 & 3.32 \\
{} & {} & {} & {} & {} & {} & {} & {} \\
1{\it b} & 1.59 $\times$ 10$^{13}$ & 2.39 $\times$ 10$^{12}$ & 0.21 & 0.10 & 
        740.2 & 1.99 $\times$ 10$^{7}$ & 6.00 & 0.60 \\
2{\it b} & 1.80 $\times$ 10$^{12}$ & 2.70 $\times$ 10$^{11}$ & 0.80 & 0.10 & 
        358.1 & 4.65 $\times$ 10$^{6}$ & 7.32 & 0.86 \\
3{\it b} & 5.25 $\times$ 10$^{11}$ & 7.87 $\times$ 10$^{10}$ & 1.00 & 0.10 & 
        237.5 & 2.05 $\times$ 10$^{6}$ & 8.44 & 1.17 \\
4{\it b} & 2.60 $\times$ 10$^{11}$ & 3.90 $\times$ 10$^{10}$ & 1.00 & 0.10 & 
        187.9 & 1.28 $\times$ 10$^{6}$ & 8.68 & 1.56 \\
5{\it b} & 1.70 $\times$ 10$^{11}$ & 2.55 $\times$ 10$^{10}$ & 1.00 & 0.10 & 
        163.1 & 9.65 $\times$ 10$^{5}$ & 8.82 & 1.86 \\
6{\it b} & 1.05 $\times$ 10$^{11}$ & 1.57 $\times$ 10$^{10}$ & 1.00 & 0.10 & 
        138.9 & 7.00 $\times$ 10$^{5}$ & 9.00 & 2.09 \\
7{\it b} & 6.13 $\times$ 10$^{10}$ & 9.19 $\times$ 10$^{9}$ & 1.00 & 0.10 & 
        116.1 & 4.89 $\times$ 10$^{5}$ & 9.21 & 2.47 \\
8{\it b} & 1.68 $\times$ 10$^{10}$ & 2.52 $\times$ 10$^{9}$ & 1.00 & 0.10 & 
        75.4 & 2.06 $\times$ 10$^{5}$ & 9.75 & 3.32 \\
{} & {} & {} & {} & {} & {} & {} & {} \\
1{\it c} & 1.50 $\times$ 10$^{13}$ & 2.25 $\times$ 10$^{12}$ & 0.22 & 0.10 & 
        726.0 & 1.91 $\times$ 10$^{7}$ & 6.07 & 0.60 \\
2{\it c} & 2.00 $\times$ 10$^{12}$ & 3.00 $\times$ 10$^{11}$ & 0.77 & 0.10 & 
        370.9 & 4.99 $\times$ 10$^{6}$ & 7.27 & 0.86 \\
3{\it c} & 6.45 $\times$ 10$^{11}$ & 9.67 $\times$ 10$^{10}$ & 1.00 & 0.10 & 
        254.4 & 2.35 $\times$ 10$^{6}$ & 8.34 & 1.17 \\
4{\it c} & 3.28 $\times$ 10$^{11}$ & 4.92 $\times$ 10$^{10}$ & 1.00 & 0.10 & 
        203.0 & 1.50 $\times$ 10$^{6}$ & 8.60 & 1.56 \\
5{\it c} & 2.22 $\times$ 10$^{11}$ & 3.33 $\times$ 10$^{10}$ & 1.00 & 0.10 & 
        178.3 & 1.15 $\times$ 10$^{6}$ & 8.73 & 1.86 \\
6{\it c} & 1.42 $\times$ 10$^{11}$ & 2.12 $\times$ 10$^{10}$ & 1.00 & 0.10 & 
        153.5 & 8.55 $\times$ 10$^{5}$ & 8.89 & 2.09 \\
7{\it c} & 8.50 $\times$ 10$^{10}$ & 1.27 $\times$ 10$^{10}$ & 1.00 & 0.10 & 
        129.5 & 6.08 $\times$ 10$^{5}$ & 9.08 & 2.47 \\
8{\it c} & 2.45 $\times$ 10$^{10}$ & 3.67 $\times$ 10$^{9}$ & 1.00 & 0.10 & 
        85.5 & 2.65 $\times$ 10$^{5}$ & 9.59 & 3.32 \\
{} & {} & {} & {} & {} & {} & {} & {} \\
1{\it d} & 1.37 $\times$ 10$^{13}$ & 2.05 $\times$ 10$^{12}$ & 0.24 & 0.10 & 
        703.5 & 1.80 $\times$ 10$^{7}$ & 6.19 & 0.60 \\
2{\it d} & 1.90 $\times$ 10$^{12}$ & 2.85 $\times$ 10$^{11}$ & 0.78 & 0.10 & 
        364.6 & 4.83 $\times$ 10$^{6}$ & 7.30 & 0.86 \\
3{\it d} & 6.20 $\times$ 10$^{11}$ & 9.30 $\times$ 10$^{10}$ & 1.00 & 0.10 & 
        251.0 & 2.29 $\times$ 10$^{6}$ & 8.36 & 1.17 \\
4{\it d} & 3.16 $\times$ 10$^{11}$ & 4.74 $\times$ 10$^{10}$ & 1.00 & 0.10 & 
        200.5 & 1.46 $\times$ 10$^{6}$ & 8.61 & 1.56 \\
5{\it d} & 2.15 $\times$ 10$^{11}$ & 3.23 $\times$ 10$^{10}$ & 1.00 & 0.10 & 
        176.4 & 1.13 $\times$ 10$^{6}$ & 8.74 & 1.86 \\
6{\it d} & 1.37 $\times$ 10$^{11}$ & 2.05 $\times$ 10$^{10}$ & 1.00 & 0.10 & 
        151.7 & 8.35 $\times$ 10$^{5}$ & 8.90 & 2.09 \\
7{\it d} & 8.25 $\times$ 10$^{10}$ & 1.24 $\times$ 10$^{10}$ & 1.00 & 0.10 & 
        128.2 & 5.96 $\times$ 10$^{5}$ & 9.09 & 2.47 \\
8{\it d} & 2.39 $\times$ 10$^{10}$ & 3.59 $\times$ 10$^{9}$ & 1.00 & 0.10 & 
        84.9 & 2.61 $\times$ 10$^{5}$ & 9.60 & 3.32 \\
\hline
\end{tabular}
\end{minipage}
\end{table*}

\subsection{Stellar nucleosynthesis prescriptions}

We compute the abundance evolution of the following elemental species: H, D, 
$^3$He, $^4$He, $^{12}$C, $^{13}$C, $^{14}$N, $^{16}$O, Ne, Mg, Si, Fe, and 
neutron-rich isotopes synthesized from $^{12}$C, $^{13}$C, $^{14}$N, and 
$^{16}$O. Starting from an initial primordial chemical composition of 24 per 
cent  $^4$He and the remaining hydrogen, the evolution of each element is 
followed by using the formalism of the production matrix $Q_{ij}(M)$, first 
introduced by Talbot \& Arnett (1973), that gives the fraction of the mass of 
an element $j$ initially present in a star of mass $M$ that is transformed in 
element $i$ and ejected. Detailed nucleosynthesis prescriptions are from: {\it 
i)} van den Hoek \& Groenewegen (1997) for low- and intermediate-mass stars 
(0.8\,--\,8 $M_\odot$) (their case with variable mass loss scaling parameter 
$\eta_{AGB}$ for stars on the asymptotic giant branch); {\it ii)} Charbonnel 
\& do Nascimento (1998) for $^3$He production/destruction in low-mass stars 
($M$ $<$ 2 $M_\odot$); {\it iii)} Nomoto et al. (1997) for Type II SNe ($M$ 
$>$ 13 $M_\odot$); {\it iv)} Thielemann, Nomoto \& Hashimoto (1993) for Type 
Ia SNe (exploding white dwarfs in binary systems). 

There has been some claim in the literature that in order to compare 
theoretical results with observations one should consider {\it all isotopes} 
rather than only the dominant ones as usually done in chemical evolution 
studies (see, e.g., Gibson \& Woolaston 1998). In Fig.\,1 we compare 
the magnesium masses ejected in form of the main isotope with those ejected 
in form of all isotopes, as a function of the initial mass of the progenitor 
star. In Fig.\,2 we do the same for iron. The same quantities are listed in 
Table 1. Notice that values for a 100 $M_\odot$ star have been extrapolated. 
As far as iron is concerned, it is nearly irrelevant to add the less abundant 
isotopes to the main one, whereas this is {\it not true} for magnesium. We 
computed both models including all isotopes and models including only the 
main ones.

\subsection{The average metallicity of a composite stellar population}

The average metallicity (or abundance, in general) which should be compared 
with the indices should be averaged on the visual light, namely:
\begin{equation}
\langle X_i \rangle_{light} = \frac{\sum_{ij} n_{ij} X_i L_{V_j}}{\sum_{ij} 
n_{ij} L_{V_j}}
\end{equation}
where $n_{ij}$ is the number of stars in the abundance interval $X_i$ and 
luminosity $L_{V_j}$. On the other hand, the real average abundance should 
be the mass-averaged one, namely:
\begin{equation}
\langle X_i \rangle_{mass} = \frac{1}{S_{tot}} \int_{0}^{S_{tot}} X_i(S) 
{\mathrm{d}}S
\end{equation}
where $S_{tot}$ is the total mass of stars ever born (Pagel \& Patchett 1975). 
Here we will mostly use this last equation, since for massive elliptical 
galaxies the difference between the mass-averaged metallicity and the 
luminosity-averaged one is almost negligible (Yoshii \& Arimoto 1987; Gibson 
1997).

\begin{table*}
\centering
\begin{minipage}{15.5cm}
\caption{Model results (case $z_f$ = 5). $M_\star(t_{QSO})$ is the stellar 
	 mass that is formed until $t_{QSO}$ is reached, expressed in units of 
	 $M_\odot$. The correction factors 1.6 (Salpeter IMF) and 2.4 (two 
	 slope IMF) should be applied in order to obtain the mass in stars 
	 dead and alive at the present time ($M_{sph}$).The chemical 
	 properties of the composite stellar population are averaged on the 
	 mass. The metallicity indices Mg$_2$ and $\langle$Fe$\rangle$ given 
	 in the last two columns are computed following the prescriptions of 
	 Matteucci et al. (1998). See text for more details.}
\begin{tabular}{@{}ccccccccccc}
\hline
Model & $M_\star(t_{QSO})$ & $\langle$[Fe/H]$\rangle$ & 
$\langle$[Mg/H]$\rangle$ & $\langle$[Mg/Fe]$\rangle$ & $\langle$[E/H]$\rangle$ 
& $\langle$[E/Fe]$\rangle$ & $\langle$[Z/H]$\rangle$ & $\langle$Z$\rangle$ & 
Mg$_2$ & $\langle$Fe$\rangle$ \\
\hline
1{\it a} & 2.92 $\times$ 10$^{11}$ & $-$0.420 & $-$0.038 & 0.382 & $-$0.113 & 
0.307 & $-$0.103 & 0.0137 & 0.232 & 2.563 \\
2{\it a} & 1.24 $\times$ 10$^{11}$ & $-$0.343 & $-$0.018 & 0.325 & $-$0.083 & 
0.260 & $-$0.072 & 0.0147 & 0.238 & 2.656 \\
3{\it a} & 5.94 $\times$ 10$^{10}$ & $-$0.345 & $-$0.084 & 0.261 & $-$0.140 & 
0.205 & $-$0.126 & 0.0131 & 0.226 & 2.624 \\
4{\it a} & 2.44 $\times$ 10$^{10}$ & $-$0.396 & $-$0.177 & 0.219 & $-$0.226 & 
0.170 & $-$0.210 & 0.0108 & 0.209 & 2.525 \\
5{\it a} & 1.40 $\times$ 10$^{10}$ & $-$0.442 & $-$0.245 & 0.197 & $-$0.290 & 
0.151 & $-$0.274 & 0.0094 & 0.198 & 2.441 \\
6{\it a} & 7.21 $\times$ 10$^{9}$ & $-$0.517 & $-$0.333 & 0.184 & $-$0.377 & 
0.140 & $-$0.359 & 0.0077 & 0.182 & 2.315 \\
7{\it a} & 3.34 $\times$ 10$^{9}$ & $-$0.609 & $-$0.443 & 0.166 & $-$0.484 & 
0.125 & $-$0.466 & 0.0061 & 0.164 & 2.158 \\
8{\it a} & 4.61 $\times$ 10$^{8}$ & $-$0.888 & $-$0.750 & 0.138 & $-$0.787 & 
0.101 & $-$0.767 & 0.0031 & 0.112 & 1.693 \\
{} & {} & {} & {} & {} & {} & {} & {} & {} \\
1{\it b} & 2.97 $\times$ 10$^{11}$ & $-$0.413 & $-$0.031 & 0.382 & $-$0.105 & 
0.308 & $-$0.096 & 0.0140 & 0.234 & 2.574 \\
2{\it b} & 1.24 $\times$ 10$^{11}$ & $-$0.317 & 0.010 & 0.327 & $-$0.056 & 
0.261 & $-$0.045 & 0.0156 & 0.243 & 2.697 \\
3{\it b} & 5.90 $\times$ 10$^{10}$ & $-$0.293 & $-$0.029 & 0.263 & $-$0.085 & 
0.207 & $-$0.072 & 0.0147 & 0.236 & 2.707 \\
4{\it b} & 2.49 $\times$ 10$^{10}$ & $-$0.326 & $-$0.106 & 0.221 & $-$0.155 & 
0.171 & $-$0.140 & 0.0127 & 0.222 & 2.635 \\
5{\it b} & 1.42 $\times$ 10$^{10}$ & $-$0.365 & $-$0.167 & 0.198 & $-$0.213 & 
0.152 & $-$0.197 & 0.0111 & 0.211 & 2.563 \\
6{\it b} & 7.27 $\times$ 10$^{9}$ & $-$0.434 & $-$0.250 & 0.184 & $-$0.294 & 
0.141 & $-$0.276 & 0.0093 & 0.197 & 2.447 \\
7{\it b} & 3.34 $\times$ 10$^{9}$ & $-$0.521 & $-$0.355 & 0.166 & $-$0.396 & 
0.125 & $-$0.378 & 0.0074 & 0.179 & 2.299 \\
8{\it b} & 4.62 $\times$ 10$^{8}$ & $-$0.791 & $-$0.653 & 0.138 & $-$0.689 & 
0.101 & $-$0.669 & 0.0038 & 0.129 & 1.851 \\
{} & {} & {} & {} & {} & {} & {} & {} & {} \\
1{\it c} & 3.04 $\times$ 10$^{11}$ & $-$0.195 & 0.288 & 0.483 & 0.209 & 
0.405 & 0.215 & 0.0275 & 0.297 & 2.949 \\
2{\it c} & 1.26 $\times$ 10$^{11}$ & $-$0.153 & 0.289 & 0.442 & 0.215 & 
0.369 & 0.222 & 0.0279 & 0.297 & 2.996 \\
3{\it c} & 5.95 $\times$ 10$^{10}$ & $-$0.201 & 0.197 & 0.398 & 0.129 & 
0.330 & 0.137 & 0.0233 & 0.279 & 2.906 \\
4{\it c} & 2.48 $\times$ 10$^{10}$ & $-$0.275 & 0.089 & 0.364 & 0.026 & 
0.301 & 0.036 & 0.0187 & 0.258 & 2.779 \\
5{\it c} & 1.42 $\times$ 10$^{10}$ & $-$0.335 & 0.013 & 0.348 & $-$0.048 & 
0.287 & $-$0.037 & 0.0159 & 0.243 & 2.679 \\
6{\it c} & 7.25 $\times$ 10$^{9}$ & $-$0.423 & $-$0.084 & 0.339 & $-$0.144 & 
0.279 & $-$0.133 & 0.0129 & 0.225 & 2.539 \\
7{\it c} & 3.32 $\times$ 10$^{9}$ & $-$0.528 & $-$0.202 & 0.326 & $-$0.261 & 
0.267 & $-$0.249 & 0.0099 & 0.202 & 2.368 \\
8{\it c} & 4.62 $\times$ 10$^{8}$ & $-$0.825 & $-$0.520 & 0.305 & $-$0.576 & 
0.249 & $-$0.563 & 0.0049 & 0.143 & 1.895 \\
{} & {} & {} & {} & {} & {} & {} & {} & {} \\
1{\it d} & 3.04 $\times$ 10$^{11}$ & $-$0.132 & 0.327 & 0.459 & 0.250 & 
0.383 & 0.256 & 0.0299 & 0.305 & 3.034 \\
2{\it d} & 1.26 $\times$ 10$^{11}$ & $-$0.093 & 0.329 & 0.423 & 0.260 & 
0.353 & 0.266 & 0.0305 & 0.306 & 3.080 \\
3{\it d} & 5.98 $\times$ 10$^{10}$ & $-$0.143 & 0.240 & 0.383 & 0.177 & 
0.320 & 0.185 & 0.0257 & 0.288 & 2.989 \\
4{\it d} & 2.48 $\times$ 10$^{10}$ & $-$0.220 & 0.132 & 0.353 & 0.075 & 
0.295 & 0.084 & 0.0207 & 0.266 & 2.85 \\
5{\it d} & 1.42 $\times$ 10$^{10}$ & $-$0.282 & 0.056 & 0.338 & 0.001 & 
0.283 & 0.011 & 0.0177 & 0.252 & 2.756 \\
6{\it d} & 7.23 $\times$ 10$^{9}$ & $-$0.372 & $-$0.044 & 0.329 & $-$0.097 & 
0.275 & $-$0.086 & 0.0143 & 0.233 & 2.612 \\
7{\it d} & 3.32 $\times$ 10$^{9}$ & $-$0.479 & $-$0.163 & 0.317 & $-$0.214 & 
0.265 & $-$0.203 & 0.0110 & 0.210 & 2.439 \\
8{\it d} & 4.64 $\times$ 10$^{8}$ & $-$0.779 & $-$0.482 & 0.297 & $-$0.530 & 
0.248 & $-$0.518 & 0.0054 & 0.151 & 1.962 \\
\hline
\end{tabular}
\end{minipage}
\end{table*}

\begin{figure*}
\begin{minipage}{15.5cm}
\centerline{\psfig{figure=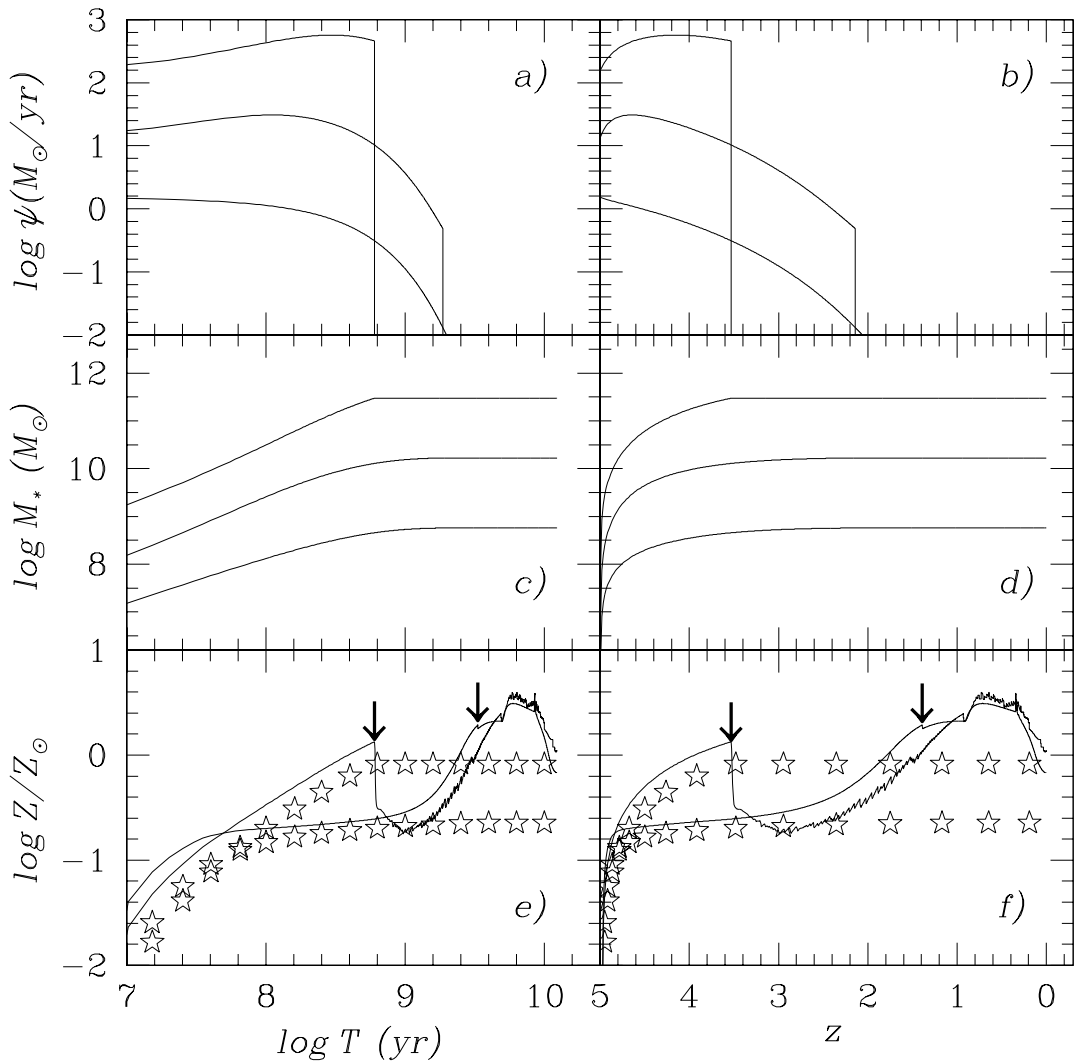,width=15.cm,height=15.cm} }
\caption{Star formation rate, cumulative mass in stars, metal enrichment of 
gas (lines) and stellar component (stars) as functions of time from the 
beginning of galaxy formation (left panels) and redshift (right panels) for 
Models 8{\it b}, 5{\it b}, and 1{\it b} ($M_{sph} \sim 3 \times 10^8 M_\odot$, 
$10^{10} M_\odot$, and $2 \times 10^{11} M_\odot$, respectively, from bottom 
to top in each panel, except for panels {\it e)} and {\it f)} where only 
results for Models 8{\it b} and 1{\it b} are shown). The arrows mark the QSO 
shining time/redshift. The adopted IMF is the Salpeter one.}
\end{minipage}
\end{figure*}

\begin{table*}
\centering
\begin{minipage}{15.5cm}
\caption{Same as Table 3 in the case of $z_f$ = 9, for set {\it d}.}
\begin{tabular}{@{}ccccccccccc}
\hline
Model & $M_\star(t_{QSO})$ & $\langle$[Fe/H]$\rangle$ & 
$\langle$[Mg/H]$\rangle$ & $\langle$[Mg/Fe]$\rangle$ & $\langle$[E/H]$\rangle$ 
& $\langle$[E/Fe]$\rangle$ & $\langle$[Z/H]$\rangle$ & $\langle$Z$\rangle$ & 
Mg$_2$ & $\langle$Fe$\rangle$ \\
\hline
1{\it d} & 3.05 $\times$ 10$^{11}$ & $-$0.136 & 0.345 & 0.481 & 0.265 & 
0.401 & 0.269 & 0.0308 & 0.309 & 3.037 \\
2{\it d} & 1.26 $\times$ 10$^{11}$ & $-$0.025 & 0.396 & 0.421 & 0.327 & 
0.352 & 0.333 & 0.0351 & 0.319 & 3.183 \\
3{\it d} & 5.96 $\times$ 10$^{10}$ & $-$0.032 & 0.348 & 0.380 & 0.287 & 
0.319 & 0.294 & 0.0324 & 0.309 & 3.159 \\
4{\it d} & 2.49 $\times$ 10$^{10}$ & $-$0.100 & 0.255 & 0.355 & 0.198 & 
0.298 & 0.207 & 0.0269 & 0.291 & 3.045 \\
5{\it d} & 1.43 $\times$ 10$^{10}$ & $-$0.156 & 0.187 & 0.343 & 0.132 & 
0.288 & 0.142 & 0.0234 & 0.277 & 2.953 \\
6{\it d} & 7.24 $\times$ 10$^{9}$ & $-$0.241 & 0.095 & 0.335 & 0.042 & 
0.283 & 0.052 & 0.0193 & 0.259 & 2.819 \\
7{\it d} & 3.32 $\times$ 10$^{9}$ & $-$0.344 & $-$0.018 & 0.325 & $-$0.069 & 
0.274 & $-$0.059 & 0.0151 & 0.237 & 2.655 \\
8{\it d} & 4.64 $\times$ 10$^{8}$ & $-$0.636 & $-$0.327 & 0.309 & $-$0.376 & 
0.261 & $-$0.364 & 0.0077 & 0.179 & 2.191 \\
\hline
\end{tabular}
\end{minipage}
\end{table*}

\section{Results}

\subsection{The chemistry of the gas and the stars}

In the reference case $\Omega_\Lambda$ = 0.7, $\Omega_M$ = 0.3, $h$ = 0.7, 
$z_{f}$ = 5, four sets of eight models each have been computed (it is worth 
noticing that $z_{f}$ can span a large redshift interval, the corresponding 
volume density of virialized haloes being specified by the Press-Schechter 
formalism -- Press \& Schechter 1974; Sheth \& Tormen 1999). The model 
parameters are shown in Table 2. The first two sets (labelled {\it a} and {\it 
b}) are relevant to a Salpeter IMF, $\phi(M)$ $\propto$ $M^{-1.35}$. Set {\it 
c} assumes a flatter IMF, $\phi(M)$ $\propto$ $M^{-1.15}$, and set {\it d} 
refers to a two slope IMF, $\phi(M)$ $\propto$ $M^{-0.4}$ for $M$ $\le$ 1 
$M_\odot$ and $\phi(M)$ $\propto$ $M^{-1.25}$ for $M$ $>$ 1 $M_\odot$. Set 
{\it a} refers to the case in which the efficiency of reheating from SNeII, 
$\epsilon$, is 15 per cent, whereas sets {\it b}, {\it c}, and {\it d} refer 
to a case with a slightly lower efficiency of reheating from SNeII, namely 10 
per cent.

Before discussing model results, we want briefly to focus on the behaviour 
of the star formation efficiency, $\nu_\star$, as a function of the galactic 
mass. $\nu_\star$ has been computed according to Eq.(11). $\nu_\star$ turns 
out to be a decreasing function of the galactic mass (see Table 2, column 8). 
Nevertheless, our star formation efficiencies yield average star formation 
rates increasing with increasing the galactic mass -- $\langle$SFR$\rangle$ 
$\propto$ $M_{sph}^{1.3}$ -- and guarantee the correct trend of increasing 
[Mg/Fe] ratios with increasing total galactic mass (see Table 3). This latter 
result is mainly due to the implementation in Eq.(9) of a reheating term 
proportional to the rate of star formation, which prevents smaller galaxies 
from efficiently converting gas into stellar generations at too early times. 
Owing to that, in smaller galaxies a significant fraction of stars form later 
on, out of gas significantly depleted in Type II SN products and enriched in 
iron by Type Ia SN explosions. Conversely, the short duration of the burst 
secures a lower relative abundance of iron with respect to magnesium in larger 
galaxies.

\begin{figure*}
\begin{minipage}{15.5cm}
\centerline{\psfig{figure=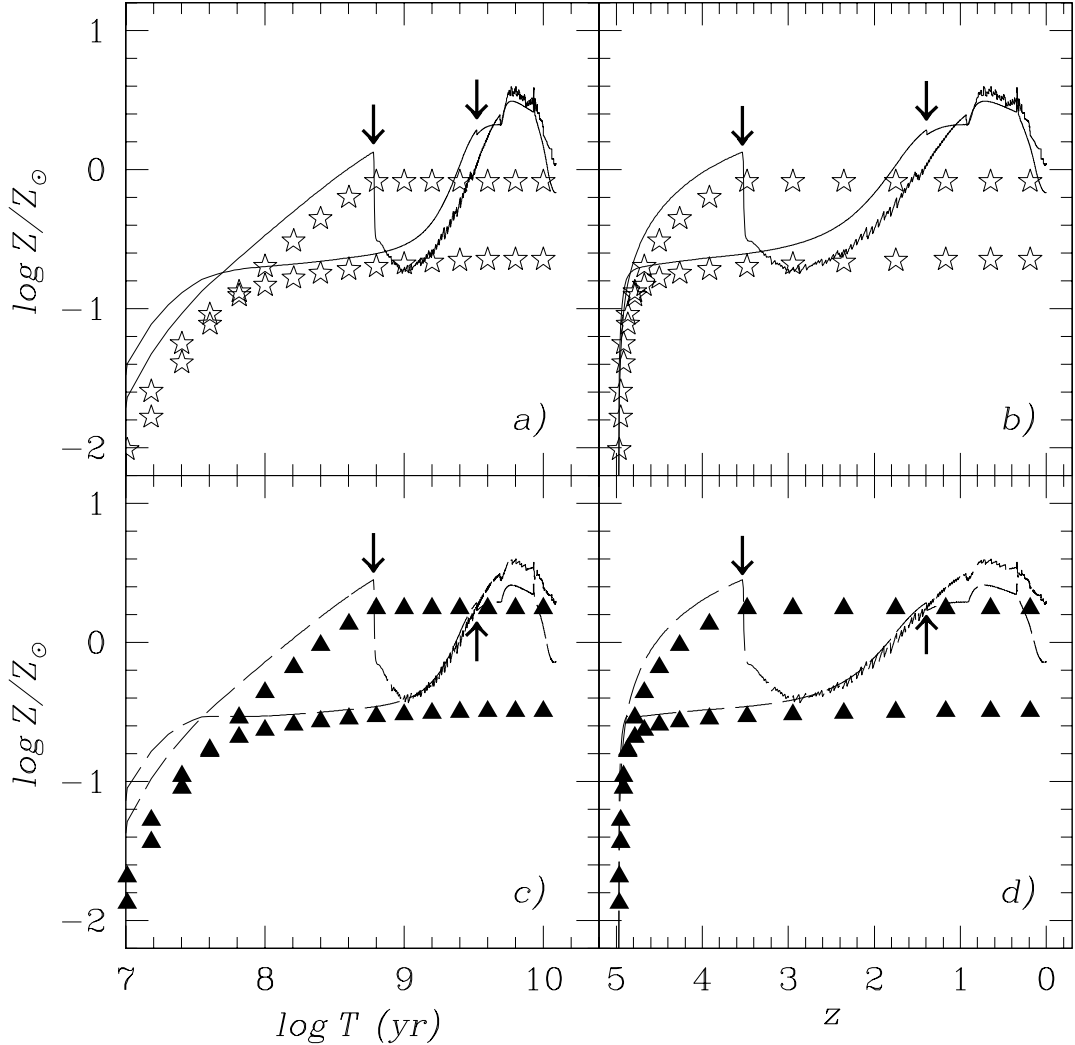,width=15.cm,height=15.cm} }
\caption{Metal enrichment of gas (lines) and stars (symbols) as functions of 
time from the beginning of galaxy formation (left panels) and redshift (right 
panels) for Models 8{\it b} ($M_{sph} \sim 3 \times 10^8 M_\odot$) and 1{\it 
b} ($M_{sph} \sim 2 \times 10^{11} M_\odot$) [from bottom to top, panels {\it 
a)} and {\it b)}] and Models 8{\it d} ($M_{sph} \sim 2 \times 10^8 M_\odot$) 
and 1{\it d} ($M_{sph} \sim 1.5 \times 10^{11} M_\odot$) [from bottom to top, 
panels {\it c)} and {\it d)}]. Set {\it b} refers to a Salpeter IMF, $\phi(M)$ 
$\propto$ $M^{-1.35}$, set {\it d} to a two slope one of the kind: $\phi(M)$ 
$\propto$ $M^{-0.4}$ for $M$ $\le$ 1 $M_\odot$ and $\phi(M)$ $\propto$ 
$M^{-1.25}$ for $M$ $>$ 1 $M_\odot$. The arrows indicate the QSO shining 
time/redshift ($t_{QSO}-t_f$ = 0.6 Gyr for Models 1{\it b,d}; $t_{QSO}-t_f$ = 
3.32 Gyr for Models 8{\it b,d}).}
\end{minipage}
\end{figure*}

In Fig.\,3 we show the star formation rate, the stellar mass and the metal 
enrichment of both the ISM and the stellar component as functions of time and 
redshift for some selected models. The models in the figure refer to present 
day galaxies with masses ranging from $\sim 3 \times 10^8 M_\odot$ to $\sim 2 
\times 10^{11} M_\odot$, and to the Salpeter IMF case. The formation of 
early-type galaxies proceeds as a maximum intensity starburst (see also 
Elmegreen 1999), reaching a rate as high as $\sim$ 1000 $M_\odot$ yr$^{-1}$ in 
the case of the most massive spheroids [upper lines in panels {\it a)} and 
{\it b)} of Fig.\,3], until the QSO shines at the centre. At that time, the 
star formation ceases. The chemical properties of the stellar populations, 
listed in Table 3 for the enlarged grid of masses, have been computed at the 
present time, assuming passive evolution since $t$ = $t_{QSO}$. It can be seen 
that, while by adopting a Salpeter IMF one can hardly attain in the gas the 
super-solar metallicity inferred from QSO spectra, flattening the IMF is an 
easy way to overcome this problem (see also Matteucci \& Padovani 1993). 
Results relevant to sets {\it b} (Salpeter IMF) and {\it d} (two slope IMF) 
are displayed in Fig.\,4: at the time of the shining of the QSO (indicated by 
the arrow), $Z_{gas}$ is $\sim$ $Z_\odot$ for Model 1{\it b} ($M_{sph} \sim 2 
\times 10^{11} M_\odot$) and $\sim$ 3 $Z_\odot$ for Model 1{\it d} ($M_{sph} 
\sim 10^{11} M_\odot$). These values are in good agreement with the range 
inferred from the analysis of the most robust diagnostics in QSO spectra, such 
as \hbox{N\,{\small III}]}/\hbox{O\,{\small III}]} and \hbox{N\,{\small 
V}}/(\hbox{C\,{\small IV}}+\hbox{O\,{\small VI}}) (Hamann et al. 2001). 
The actual QSO metallicities could be as much as 2\,--\,3 times higher than 
the above estimates (Hamann et al. 2001; Warner et al. 2001). However, it 
should be stressed that observed values refer to the BELRs and reflect the 
composition of the very central region of the host galaxy. Our predictions, 
derived with a one-zone chemical evolution model, are indeed lower limits for 
the central regions.

As far as the low-luminosity galaxy is concerned (Models 8 {\it b,d}; $M_{sph} 
\sim 2$\,--\,$3 \times 10^8 M_\odot$), negligible differences are found by 
changing the IMF slope. This is due to the role of the stellar feedback in 
low-mass galaxies, which becomes increasingly important when favoring massive 
stars with respect to low- and intermediate-mass stars in the IMF: whereas in 
high-mass spheroids the formation of the stellar bulge is ruled by the cooling 
of the gas and the shining of the QSO, in low-mass spheroids the building up 
of the stellar population is rather controlled by the feedback. 

Recent estimates suggest values of [Mg/Fe] between 0.40 and 0.00 at the centre 
of ellipticals (Kuntschner 2000; Kuntschner et al. 2001; Terlevich \& Forbes 
2001). These values are derived from {\it central} absorption-line strengths. 
Corrections for line-strength gradients have to be applied in order to get the 
integrated or composite indices, representative of the mean stellar 
metallicities, which can be compared with results from composite models 
representing {\it whole} galaxies. Kobayashi \& Arimoto (1999), by using the 
metallicity gradients of 80 elliptical galaxies and by assuming that Mg$_2$ 
and Fe$_1$ reflect the abundances of magnesium and iron, respectively, find 
$\langle$[Mg/Fe]$\rangle$ $\simeq$ +0.2 in most of early-type galaxies. Their 
$\langle$[Mg/Fe]$\rangle$ does not correlate with galaxy mass tracers, at 
variance with the central [Mg/Fe]. However, as the authors themselves stress, 
the ratio of Mg$_2$ to Fe$_1$ may not directly give the [Mg/Fe] ratio. Milone, 
Barbuy \& Schiavon (2000) give instead $\langle$[Mg/Fe]$\rangle$ $\simeq$ 
+0.3\,--\,+0.4 (see their fig.\,5{\it b}). Our theoretical 
$\langle$[Mg/Fe]$\rangle$ are listed in Table 3. They display a decreasing 
trend with decreasing galactic mass, even if milder than observed at the 
galactic centre. The $\langle$[Mg/Fe]$\rangle$ ratios listed in Table 3 have 
been computed by using the yields of $^{24}$Mg+$^{25}$Mg+$^{26}$Mg and 
$^{54}$Fe+$^{56}$Fe+$^{57}$Fe+$^{58}$Fe, rather than those of the main 
isotopes alone, following arguments by Gibson \& Woolaston (1998). The 
elemental $\langle$[$^{24}$Mg/$^{56}$Fe]$\rangle$ ratios turn out to be only 
less than 0.10 dex lower on the average. Owing to the observational 
uncertainties, we can not discriminate between the two choices.

If we consider the ratio $\langle$[E/Fe]$\rangle$ rather than 
$\langle$[Mg/Fe]$\rangle$, where E refers to all `enhanced' elements, namely 
C, N, O, Ne, Mg, Si, S following Trager et al. (2000a, their model 4 -- see 
below), we find milder enhancements. This is due to the fact that now we are 
considering also elemental species partially (Si, S) if not even almost 
entirely (C, N) produced by low- and intermediate-mass stars restoring their 
nucleosynthesis products on longer time-scales than SNeII (responsible for the 
whole Mg production). In Fig.\,5 we show the temporal evolution of the 
abundance ratios of some enhanced species with respect to iron (from $t$ = 
$t_f$ to $t$ = $t_{QSO}$) for Models 1 {\it b,d}. Different enhanced species 
behave very differently; by the way, C should not be included in the enhanced 
group at all. 

\begin{figure}
\centerline{\psfig{figure=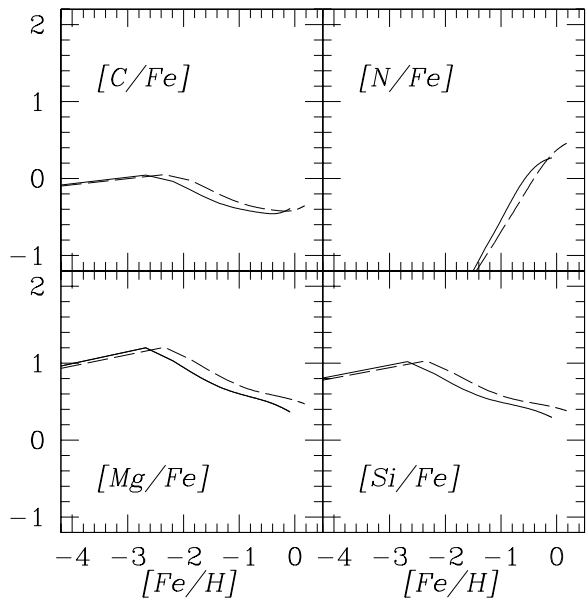,width=8.3cm,height=8.3cm} }
\caption{[C, N, Mg, Si/Fe] vs. [Fe/H] in the gas from $t$ = $t_f$ to $t$ = 
$t_{QSO}$ for Models 1{\it b} (continuous lines) and 1{\it d} (dashed lines). 
Different elements show different degrees of enhancement, owing to their 
different nucleosynthesis histories.}
\end{figure}

Kuntschner \& Davies (1998), from a comparison of their measurements of 
central C\,4668 and H$_{\gamma_{A}}$ indices in a sample of early-type 
galaxies in the Fornax cluster to predictions from single-burst stellar 
population models (Worthey 1994; Worthey \& Ottaviani 1997), derive that their 
ellipticals have [Fe/H] from $-$0.1 to $+$0.6. Kuntschner (2000) finds that 
early-type galaxies brighter than $M_B$ = $-$17 in the Fornax cluster form a 
sequence in metallicity varying roughly from $-$0.25 to $+$0.30 in central 
[Fe/H]. Making an attempt at estimating mean values, Kobayashi \& Arimoto 
(1999) find that typical mean stellar metallicities are 
$\langle$[Fe/H]$\rangle$ $\simeq$ $-$0.3 and range from 
$\langle$[Fe/H]$\rangle$ $\simeq$ $-$0.8 to +0.3, well below the highest 
values observed at the galactic centres. Our predictions (Table 3, third 
column) are indeed in good agreement with the values inferred from the 
metallicity gradients, especially if an IMF flatter than the Salpeter one is 
assumed. However, we never recover the highest metallicity values. 
Undoubtedly, the fact that in the framework of our model we can not assembly 
objects more massive than $M_{sph}$ $\sim$ 1.5\,--\,2 $\times$ 10$^{11}$ 
$M_\odot$ is playing a crucial role (however, it should be noticed that the 
maximum stellar mass which can be assembled depends also on the assumed 
redshift of galaxy formation -- see Sect.4.3 below).

Very recently, Trager et al. (2000a) have derived single-burst stellar 
population equivalent ages, metallicities, and abundance ratios for a sample 
of local early-type galaxies from H$\beta$, Mg$b$, and $\langle$Fe$\rangle$ 
line strenghts using an extension of the Worthey (1994) models that accounts 
for the enhancements of Mg and other $\alpha$-elements relative to the Fe-peak 
elements. The metallicities and enhancement ratios [E/Fe] they have found are 
strongly peaked around +0.26 and +0.20, respectively, in an aperture of radius 
$r_e$/8. Gradients in stellar populations within galaxies are found to be 
mild, with metallicity decreasing by 0.20 dex and [E/Fe] remaining nearly 
constant out to an aperture of radius $r_e$/2 for nearly all systems. In 
Fig.\,6 we compare our $\langle$[E/H]$\rangle$ vs. $\langle$[Fe/H]$\rangle$ 
relation for Models from 1{\it b,d} to 8{\it b,d} (spheroid masses range from 
$M_{sph} \sim 2 \times 10^{11} M_\odot$ to $M_{sph} \sim 2 \times 10^8 
M_\odot$; set {\it b}: Salpeter IMF, set {\it d}: two slope IMF) to that by 
Trager et al. (2000a,b). Results relevant to Models from 1{\it c} to 8{\it c} 
(single slope IMF flatter than the Salpeter) are very similar to those 
relevant to Models from 1{\it d} to 8{\it d} and therefore have not been 
plotted in the graph. As expected, our metallicities and abundance ratios 
agree much better with the Trager et al.'s ones computed through the global 
$r_e$/2 aperture rather than with those computed through the central $r_e$/8 
aperture. Again, the discrepancy is mainly due to the fact that our abundances 
are accounting for the distribution of stellar populations over the whole 
galactic physical dimension. Besides this, a remarkable point is that the 
slope of our $\langle$[E/H]$\rangle$ vs. $\langle$[Fe/H]$\rangle$ relation is 
quite similar to that found by Trager et al. (2000a,b) for central values. The 
situation is also improving if galaxy formation is pushed at higher redshift 
(Table 4).

\begin{figure}
\centerline{\psfig{figure=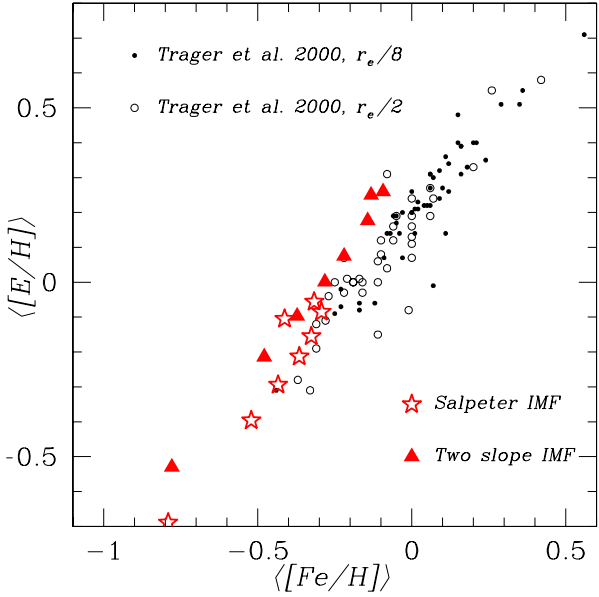,width=8.3cm,height=8.3cm} }
\caption{The $\langle$[E/H]$\rangle$ vs. $\langle$[Fe/H]$\rangle$ relations 
obtained for Models from 1{\it b,d} to 8{\it b,d} ($M_{sph} \sim 2 \times 
10^{11}$\,--\,$2 \times 10^8 M_\odot$; set {\it b}: Salpeter IMF, set {\it d}: 
two slope IMF) are compared to that found by Trager et al. (2000a,b) from 
H$\beta$, Mg$b$, and $\langle$Fe$\rangle$ measurements in a sample of local 
ellipticals. Our models show a lower Fe and E content, owing to the fact that 
we are running one zone models. Results relevant to Models from 1{\it c} to 
8{\it c} (single slope IMF flatter than the Salpeter one) are very similar to 
those relevant to Models from 1{\it d} to 8{\it d} and are therefore not shown 
here.}
\end{figure}

\begin{figure}
\centerline{\psfig{figure=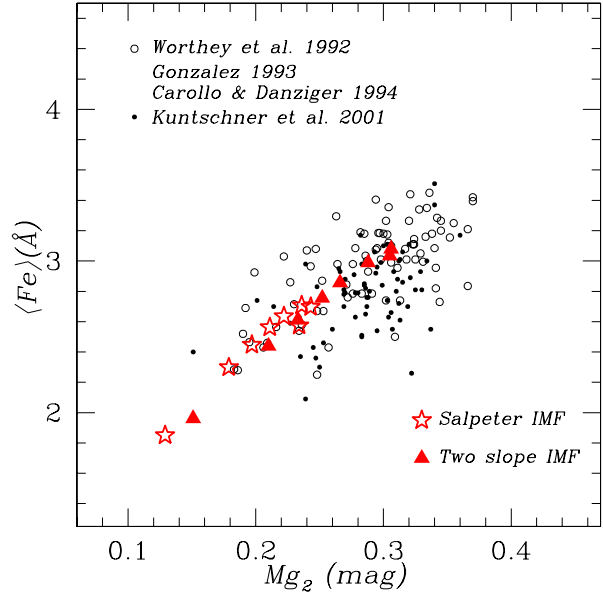,width=8.3cm,height=8.3cm} }
\caption{Mg$_2$ vs. $\langle$Fe$\rangle$ theoretical relations compared to the 
available data. The recent, homogeneous sample by Kuntschner et al. (2001) is 
shown as filled circles; older data are shown as open circles. Note that the 
Kuntschner et al. sample refers mostly to group and cluster environments. 
Results for sets {\it b} (stars) and {\it d} (triangles) are shown. The two 
sets differ uniquely in the IMF slope. Results relevant to set {\it c}, 
characterized by a single slope IMF flatter than the Salpeter one, are very 
similar to those relevant to set {\it d}, so we do not show them in the graph.}
\end{figure}

So far, we made an attempt to compare theoretical abundances for integrated 
early-type galaxies with estimates coming from the analysis of observed  
spectral indices. Now viceversa we want to use some calibrations and translate 
the theoretical abundance ratios into metallicity indices to be directly 
compared with observations. When dealing with early-type galaxies, the most 
widely used tool to compare theoretical predictions on $\alpha$-element and Fe 
abundances in stars to observations is the Mg$_2$ vs. $\langle$Fe$\rangle$ 
diagram. In Fig.\,7 observations (Worthey et al. 1992; Gonz\'alez 1993; 
Carollo \& Danziger 1994a,b; Kuntschner et al. 2001) are shown as long as 
model predictions. The theoretical indices have been obtained by converting 
the $\langle$[Mg/Fe]$\rangle$ and $\langle$[Fe/H]$\rangle$ ratios predicted by 
Models from 1{\it b,d} to 8{\it b,d} listed in Table 3 into the metallicity 
indices Mg$_2$ and $\langle$Fe$\rangle$ by means of the calibrations by 
Tantalo et al. (1998), as described in Matteucci, Ponzone \& Gibson (1998). 
The spread in the Mg$_2$ data is reproduced remarkably well. In particular, 
the adoption of a two slope IMF of the kind of that described in Sect.3.1 
allows us to cover a wider range of values in both indices. This is an 
unprecedent result, particularly interesting owing to the fact that an IMF of 
the kind adopted here is supported both observationally in the low-mass range 
(e.g., Wyse et al. 1999; Zoccali et al. 2000) and theoretically (Elmegreen 
2000a,b). Note that an IMF with a single slope of 1.15 over the entire mass 
range would produce similar results. The high-mass end of the diagram, which 
is actually missed by our theoretical models, could be easily covered by 
adopting an Arimoto \& Yoshii IMF, $\phi(M)$ $\propto$ $M^{-0.95}$, i.e., an 
IMF with an even flatter slope. However, as we already noted before, an offset 
between model predictions and data has to be expected, since data refer to 
central index values, whereas model predictions are relevant to the whole 
physical dimension of the galaxy. Milone et al. (2000) find that the 
integrated indexes of their sample galaxies span a range from $\sim$ 0.2 to 
$\sim$ 0.3 in Mg$_2$, and from $\sim$ 2 to $\sim$ 2.5 in $\langle$Fe$\rangle$, 
while central index values range from $\sim$ 0.25 to $\sim$ 0.35 for Mg$_2$, 
and from $\sim$ 2.5 to $\sim$ 3.5 for $\langle$Fe$\rangle$. The offset we find 
is nearly the same (cfr. model predictions with data in Fig.\,7). Moreover, 
as we already pointed out, in a monolithic picture for galaxy formation a 
natural cut-off in the stellar mass of galaxies should be expected, owing to 
the increase of the cooling time with increasing the host halo mass.

Finally, one could inquire about the properties of the very rare ellipticals 
formed at the highest redshifts, which are probably the hosts of the highest 
redshift QSOs ($z$ $\ga$ 6). In Table 4 we summarize the results relevant to 
set {\it d}, for the case $z_f$ = 9 and $T_{b}^*$ = 0.4 Gyr.

\subsection{Broad-band photometry}

We have checked that also the photometric properties of our elliptical galaxy 
models are consistent with observations of local galaxies. To this aim, we 
have used the spectrophotometric code by Silva et al. (1998) and computed the 
spectral energy distribution (SED) and colours expected according to the star 
formation and chemical evolution histories of the models.

For the shape of the SED, an example is shown in Fig.\,8, where we compare 
with the template SED for ellipticals by Arimoto (1996).

As discussed in Section 2.1, elliptical galaxies are observed to follow a CMR, 
with redder colours at increasing luminosity. The slope of this relation is 
interpreted as driven mostly by metallicity, rather than age, and its 
tightness as a small age dispersion among the bulk of the stellar populations. 
This interpretation rests on the reproduction, by several authors, of a 
combined set of observational constraints, including broad-band magnitudes and 
colours, spectral indexes, such as H$\beta$, Mg$_2$, $\langle$Fe$\rangle$, 
etc., and abundance ratios (e.g., Bower et al. 1992; Bressan et al. 
1994, 1996; Tantalo et al. 1996; Maraston \& Thomas 2000). On the other hand, 
there are evidences, from this same data set for local galaxies and from 
higher-$z$ observations, that at least a minor fraction of the stellar 
populations of elliptical galaxies may have formed someway after the bulk of 
their stars was already formed (e.g., Bressan et al. 1996; Franceschini et al. 
1998; Longhetti et al. 2000; Trager et al. 2000a,b).

\begin{figure}
\centerline{\psfig{figure=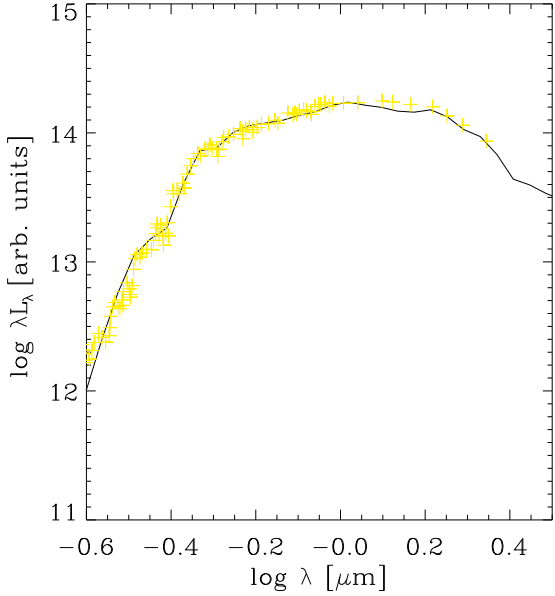,width=8.7cm,height=8.7cm} }
\caption{The SED of Model 1$d$ ($M_{sph} \sim 1.5 \times 10^{11} M_\odot$, two 
slope IMF) at 13 Gyr is compared with the template optical-NIR SED for 
elliptical galaxies by Arimoto (1996).}
\end{figure}

In Fig.\,9, the colour\,--\,magnitude relation $V-K$ vs. $M_V$ and the $J-K$ 
vs. $V-K$ colours are shown for models of case $d$ (two slope IMF, Table 3) 
with mass $M_{sph} \geq 10^{10} M_\odot$ at the age of 10 and 15 Gyr ($z_{f}$ 
= 5 implies an age of 12 Gyr for the assumed cosmological parameters). Lower 
mass models, with prolonged star formations, have more prolonged evolutionary 
histories, for which we do not have enough constraints in the framework of our 
model. The $V-K$ colour is a strong function of the global metallicity of 
stars. In agreement with the results we found for the chemical abundances, the 
colours we obtain with the Salpeter IMF are too blue because of the very low 
metallicities reached (cases $a$ and $b$ in Table 3). Instead, in order to 
reproduce the observed $V-K$ colours, an average metallicity greater than 
solar must be reached, for the most massive models, which is the case for 
Models $d$ shown in Fig.\,9. As remarked in the previous section, since the 
most massive objects in our model have $M_{sph} \simeq 1.5 \times 10^{11} 
M_\odot$, we do not cover the highest luminosity end of the plot. In Fig.\,9 
we also show the effect, for the two less massive objects in the plot at 15 
Gyr, of a small burst episode involving 5 per cent of the mass, and taking 
place 4 Gyr ago. It is clear that the slope of the predicted CMR may fluctuate 
around the average value determined by metallicity because of these 
uncertainties.

\begin{figure}
\centerline{\psfig{figure=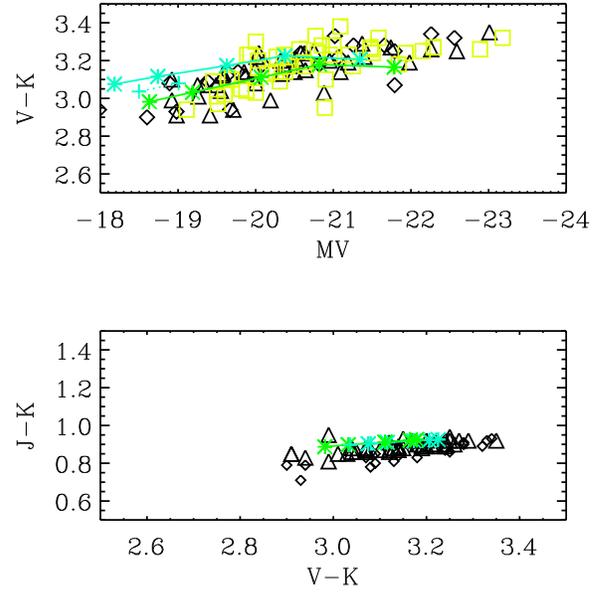,width=8.7cm,height=8.7cm} }
\caption{{\it Upper line:} The colour\,--\,magnitude relation $V-K$ vs. $M_V$ 
for Models {\it d} (two slope IMF) is shown at the age of 10 and 15 Gyr. Data 
are from Bower et al. (1992) (triangles and diamonds for Coma and Virgo 
cluster ellipticals, respectively) and Mobasher et al. (1999) (squares, Coma 
cluster ellipticals). The crosses connected by the dotted line are an example 
of how a small star formation episode may affect the $V$ magnitude (see text). 
{\it Lower panel:} $J-K$ vs. $V-K$, same models as above and data from Bower 
et al. (1992).}
\end{figure}

The spectrophotometric code provides also the blue luminosities we need in 
order to compute the rates of Type Ia SNe at the present time. We find an 
average present SNIa rate varying from $\langle Rate_{SNIa}^{th}\rangle$ = 0.3 
SNu for $z_{f}$ = 5 to $\langle Rate_{SNIa}^{th}\rangle$ = 0.2 SNu for $z_{f}$ 
= 9, to be compared with the observed value of 0.18 $\pm$ 0.06 SNu (Cappellaro 
et al. 1999). Although uncertainties in both observations (e.g., empirical 
corrections for the bias in the inner galactic regions) and theory (e.g., 
fraction of stars of a given mass that enters the formation of binary systems 
which will give rise to Type Ia SN explosions; metallicity effects -- which we 
do not account for here) prevent us from drawing any firm conclusion, we 
notice that theoretical results agree quite well with observations, at least 
at the 2\,$\sigma$ level.

\subsection{The $\bmath{M/M_{sph}}$ Ratio}

While the luminous component of a galaxy is rather well defined, the 
associated halo is not easily identifiable as a well defined structure, 
separated from other close haloes. Nevertheless, luminous galaxies are good 
tracers of the mass distribution. Recent statistical studies of the weak 
lensing associated to nearby `lens' galaxies allow for a first estimate of 
the galaxy-mass correlation function (McKay et al. 2001). These authors point 
out that the galaxy-mass correlation function is independent of the 
environment, confirming  that the mass within $\sim$ 300 $h^{-1}$ kpc from a 
central galaxy is clearly associated with the central object. They also find 
that elliptical galaxies in the luminosity range 1.6 $\leq$ $L_{z'}$ $\leq$ 
17.7 $\times$ 10$^{10}$ $h^{-2}$ $L_{\odot}$ exhibit an average $\langle 
(M(R \leq 260 \ $kpc$)/L_{z'}\rangle = 123 \pm 14 \ h$ $M_{\odot}/L_{\odot}$. 
This result holds for dark matter haloes with isothermal density profile, 
while in the case of NFW profile the average value is about a half. Assuming a 
Salpeter IMF, our spectrophotometric model yields $M/L_{z} \simeq 2.4 \ 
M_{\odot}/L_{\odot}$ for elliptical galaxies with an age $T_{gal}$ = 13 Gyr 
while, for the two slope IMF we adopted, $M/L_{z} \simeq 1.8 \ 
M_{\odot}/L_{\odot}$. In the case of a NFW profile and $H_{\circ}=70$ km 
s$^{-1}$ Mpc$^{-1}$, the results obtained by McKay et al. (2001) suggest 
$M_{halo}/M_{sph} \simeq  17$ for a Salpeter IMF and $M_{halo}/M_{sph} \simeq$ 
23 for the two slope IMF adopted in this paper, for galaxies with mass in the 
range 6 $\times$ 10$^{10}$\,--\,6 $\times$ 10$^{11} \ M_{\odot}$. From 
Fig.\,10 it is apparent that in the mass interval explored the agreement 
between the results of McKay et al. (2001) and the model predictions is very 
good.

\begin{figure}
\centerline{\psfig{figure=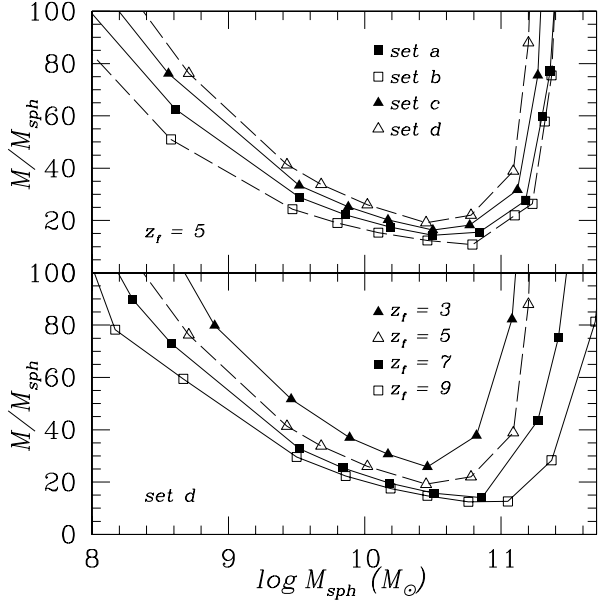,width=8.3cm,height=8.3cm} }
\caption{$M/M_{sph}$ ratios vs. $M_{sph}$. $M$ is the dark halo mass, 
$M_{sph}$ is the mass in stars and stellar remnants at the present time. The 
ratios depend on both IMF slope ({\it upper panel}; sets {\it a}, {\it b}: 
Salpeter IMF, set {\it c}: single slope IMF flatter than the Salpeter, set 
{\it d}: two slope IMF; the efficiency of reheating from SNeII is set to 10 
per cent for all cases, but set {\it a} where it is 15 per cent) and redshift 
of galaxy formation ({\it lower panel}).}
\end{figure}

The increase of the ratio  $M_{halo}/M_{sph}$ with decreasing mass in the 
low-mass range is due to the effect of the stellar feedback acting on the gas 
in shallow potential wells. This portion of the curve is sensitive to the 
fraction of the SN energy transferred to the gas, and to the IMF. As a result, 
the fraction of the baryons which form stars is strongly decreasing below a 
few 10$^{9} \ M_{\odot}$, in keeping with observations showing that in low 
luminosity/mass galaxies the relative fraction of dark matter to stars 
progressively increases with decreasing luminosity (e.g., Persic, Salucci \& 
Stel 1996).

At large masses the paroxysmal increase of the ratio with stellar mass is 
mainly due to the increase of the cooling time of the gas in the outer regions 
of large haloes. The fraction of baryons available for cooling and collapsing 
into stars is fixed by imposing $T_{burst}\leq t_{QSO}-t_{vir}$ [cfr. Eq.(8)]. 
The amount of stars formed is limited by the relatively short time lag between 
the virialization and the QSO appearance for massive objects. The 
$M_{halo}/M_{sph}$ ratio is not only sensitive to $T_{burst}$, but also to the 
formation redshift, larger masses being more easily assembled at high 
redshift, since at fixed mass the densities of DM and baryons increase with 
$z_{vir}$ (see Section 2.1). In the case of $z_{vir}$ = 9, the largest mass in 
stars in our set {\it d} with still acceptable $M_{halo}/M_{sph}$ is $M_{sph} 
\simeq 4 \times 10^{11}$ $M_\odot$ (Fig.\,10). In this framework, more 
massive galaxies are assembled by coalescence, as suggested by their often 
observed distorted morphologies (e.g., double cores). On the other hand, the 
number density of massive haloes strongly decreases with increasing redshift. 
The result is the observed exponential decline of the luminosity/mass function 
at large luminosities/masses. The statistical aspects implied by these results 
will be discussed in a forthcoming paper.

It is worth noticing that the values of $\epsilon$, namely the efficiency of 
reheating from SNeII, are significantly constrained by the results of McKay et 
al. (2001). If $\epsilon$=1 is adopted, as it seems to be required in order to 
get, under very favorable hypotheses, $\sim$ 1 KeV of energy per particle 
injected into the IGM by SNe (Valageas \& Silk 1999; Kravtsov \& Yepes 2000), 
a minimum value of $M/L_z\simeq 250$ for $L_z=2 \times 10^{10} L_{\odot}$ and 
for the case {\it d} and $z_f$ = 5 is recovered. This value exceeds by a 
factor of 4 the result of McKay et al. (2001) for a NFW density profile.

\section{Discussion and final remarks}

In this work we concentrated mostly on the chemo-photometric properties of the 
stellar populations of early-type galaxies. In particular, we tested specific 
histories for the formation of the spheroids, which turned out to provide a 
good fit to crucial observed relations: the mass\,--\,metallicity relation, 
the Mg$_2$ vs. $\langle$Fe$\rangle$ diagram, the colour\,--\,magnitude 
relation $V-K$ vs. $M_V$, the $J-K$ vs. $V-K$ colours and the 
$M_{halo}/M_{sph}$ ratio. We found that all these properties are well 
reproduced by using the same dependence of the burst duration on galactic mass 
which is requested in order to fit the submillimetre source counts (Granato et 
al. 2001). It is worth noticing that the adopted scenario does hold for 
galaxies which are large enough to host a QSO, i.e., $M_{sph}$ $\ga$ a few 
10$^{9}$ $M_\odot$, less massive objects possibly suffering for even more 
complex evolutionary paths than described above. Secondary episodes of star 
formation in massive galaxies at $z < 1$ would produce only a tiny amount of 
stars; thus, the observed evolution of the FP and the observed [O\,{\small 
II}] emission could be explained (Treu et al. 2001) without any need of 
altering the global chemical properties of the stellar population.

At variance with most of the previous models of chemical evolution of 
spheroidal galaxies, we accounted for the stellar feedback through Eq.(13), 
i.e., we allowed a fraction of gas to be heated and subtracted to the ongoing 
star formation process {\it at each time}. The reheating term acts in the 
direction of reducing the efficiency of star formation in less massive 
spheroids. This reconciles our star formation efficiencies, decreasing with 
increasing galactic mass, with the increase of the star formation efficiency 
with increasing galactic mass required by other authors and it yields 
$\langle$SFR$\rangle$ $\propto$ $M_{sph}^{1.3}$. The overall result is 
similar, none the less the physical approach is very different. In our model, 
the star formation is governed by the rate at which baryonic gas cools and 
falls into DM haloes and is inhibited by heating from SNe and definitively 
ended by the QSO peak activity combined to the SN feedback.

We find that the ratios $\langle$[Fe/H]$\rangle$, $\langle$[Mg/H]$\rangle$, 
$\langle$[Mg/Fe]$\rangle$, $\langle$[E/H]$\rangle$, $\langle$[E/Fe]$\rangle$, 
and $\langle$[Z/H]$\rangle$ {\it all increase} with galactic mass (see Table 
3). This is due to the combined effects of the following aspects of the model: 
{\it i)} the duration of star formation is shorter with increasing galactic 
mass; {\it ii)} the number of stellar generations increases with increasing 
galactic mass ($\langle$SFR$\rangle$ $\propto$ $M_{sph}^{1.3}$);  {\it iii)} 
the stellar feedback is more effective in low-mass galaxies (the amount of 
SNII-enriched material retained by low-mass galaxies is lower, due to the 
shallower potential wells).

If we relaxe the first hypothesis and set $T_{burst}$ = 2 Gyr, constant with 
mass, the abundance ratios still increase with increasing galactic mass but 
$\langle$[Mg/Fe]$\rangle$ and $\langle$[E/Fe]$\rangle$, which flatten out 
(see also Kawata 2001). Therefore, a star-forming phase which lasts longer in 
less massive spheroids is strongly needed in order to reproduce the observed 
trend of increasing [Mg/Fe] in the nuclei of local ellipticals with increasing 
galactic mass. On the contrary, if we suppress the stellar feedback 
($\epsilon$ = 0), the ratios of all the elements with respect to hydrogen, 
$\langle$[el/H]$\rangle$, decrease with increasing mass, while the increase in 
$\langle$[el/Fe]$\rangle$ is preserved.

Flattening the IMF with increasing galactic mass is a further way of 
increasing the quantities $\langle$[Fe/H]$\rangle$, $\langle$[Mg/H]$\rangle$, 
$\langle$[Mg/Fe]$\rangle$, $\langle$[E/H]$\rangle$, $\langle$[E/Fe]$\rangle$, 
and $\langle$[Z/H]$\rangle$ with increasing galactic mass; however, in the 
framework of our model, we do not need such a flattening in order to increase 
the overall mean metallicity of the stellar population.  A careful analysis of 
Tables 3 and 4 actually reveals that the mean metallicity of the stellar 
population starts slightly decreasing again at $M_{sph} \ga 10^{11} M_\odot$ 
(cfr. the el/H abundance ratios for Models 1 and 2 in both tables). This is 
due to the increase of the cooling time of the gas in the outer regions of the 
largest virialized haloes. This increase limits the amount of stars which are 
born and hence the metal enrichment of the gas.

\begin{figure}
\centerline{\psfig{figure=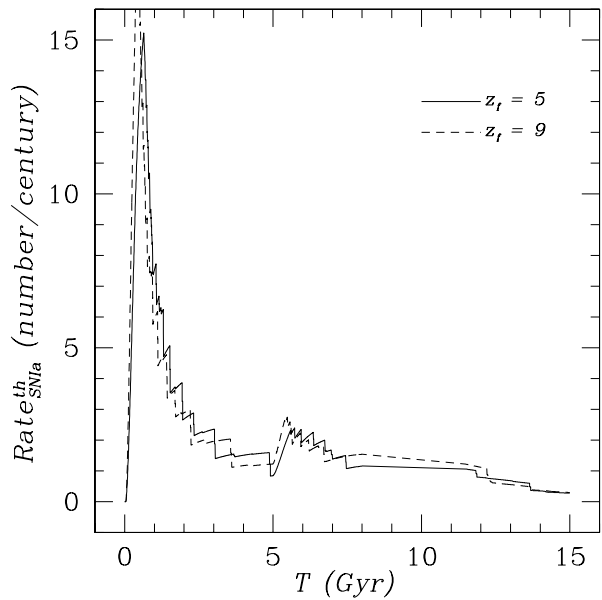,width=8.3cm,height=8.3cm} }
\caption{Theoretical Type Ia SN rate for Model 1{\it d} ($M_{sph} \sim 1.5 
\times 10^{11} M_\odot$, two slope IMF) as a function of time since the 
beginning of galaxy formation. The bulk of SNIa explosions is expected to 
occur soon after the shining of the QSO, when the galaxy shows up as an 
extremely red object (ERO).}
\end{figure}

In the framework of the adopted model, massive QSO hosts ($M_{sph} > 2 \times 
10^{10} M_\odot$) experience a relatively short ($T_{burst} \sim$ 0.5\,--\,1 
Gyr), intense (SFR $\sim$ 50\,--\,1000 $M_\odot$ yr$^{-1}$) star-forming phase 
at high redshift ($z \ga$ 3), during which they show up as ultraluminous 
far-IR galaxies ({\it SCUBA} galaxies). After the shining of the QSO, they 
appear as EROs, for which we predict solar or higher than solar stellar 
metallicities. Moreover, we predict that very high rates of SNeIa should be 
observed in EROs, since in these objects the bulk of SNeIa explode soon after 
the shining of the QSO (see Fig.\,11), in a dust-free medium.  Therefore, non 
dusty EROs are preferred targets for searching for high-$z$ SNeIa.

Intermediate- and low-mass QSO hosts ($M_{sph} < 2 \times 10^{10} M_\odot$) 
are characterized by a more protracted starburst phase ($T_{burst} \sim$ 
1\,--\,2.5 Gyr) and show up as Lyman break galaxies (LBGs) at $z \ga$ 3 (see 
also Granato et al. 2001). The curves relevant to Model 5{\it b} in the upper 
panels of Fig.\,3 are representative of the history of star formation of a 
typical LBG; more generally, our models predict SFRs between $\sim$ 5 and 
$\sim$ 50 $M_\odot$ yr$^{-1}$ for this kind of objects. These values agree 
quite well with recent estimates of the star formation activity in LBGs by 
Pettini et al. (2001). Moreover, our models predict a value of log(O/H)\,+\,12 
$\sim$ 8.2\,--\,9.2 in the gas of LBGs at $z \sim 3$, strongly depending on 
the adopted redshift of galaxy formation and hence on the age of the stellar 
population. The lowest log(O/H)\,+\,12 value we give is associated to a 
low-mass spheroid (Model 7) which started forming stars at $z_f \sim 3$ and to 
a stellar population of age $0.5 \times 10^8$ yr; the highest one is 
associated to a more massive spheroid (Model 4) which started forming stars at 
$z_f \sim 5$ and to a stellar population of age $10^9$ yr. The lowest values 
we find agree well with the estimates by Pettini et al. (2001). This is quite 
reassuring, since their estimates refer to objects with virial masses of about 
10$^{10}$ $M_\odot$, and Model 7 indeed refers to an object with virial mass 
$M_{bar} \simeq 1.2 \times 10^{10} M_\odot$ (see Table 2). A more detailed 
study of LBGs will be the subject of a forthcoming paper.

The adopted model implies that the ratio $M_{BH}/M_{sph}$ between the final 
mass of the black hole responsible for the QSO activity and the mass of the 
host galaxy is not evolving with redshift, at variance with respect to 
semi-analytical models (Kauffmann \& Haehnelt 2000). The most recent 
observations show a modest, if any, variation of this ratio with redshift 
(Kukula et al. 2001; see also the discussion in Granato et al. 2001).

The ratio $M_{halo}/M_{sph}$ predicted by the model depends significantly on 
the adopted value of $\epsilon$, the efficiency of the transfer of the SN 
energy output to the ISM. A very good agreement with the findings of McKay et 
al. (2001), $M_{halo}/M_{sph} \simeq 20$ for galaxies in the mass range 6 
$\times$ 10$^{10}$\,--\,6 $\times$ 10$^{11}$ $M_{\odot}$, is found under the 
assumption $\epsilon$ $\simeq$ 0.1. The combination of potential well and 
feedback makes the ratio increasing with decreasing mass. The assumption 
$\epsilon$=1 is strongly ruled out in our model by the fact that the minimum 
ratio $M_{halo}/M_{sph}$ would be 4 times larger than that found by McKay et 
al. (2001).

This last result is also important in the context of the preheating of the IGM 
in galaxy groups and clusters. Actually, it has been suggested that with 
$\epsilon=1$ and under very favorable conditions, SNe can yield as much as 
$\sim$ 0.5\,--\,1 KeV per particle, thus bringing theoretical models of 
cluster formation into agreement with observations. However, this would 
contrast with the $M_{halo}/M_{sph}$ ratio found in galaxies. The supernovae 
are likely to be supplemented by some other heating mechanism (Balogh, Babul 
\& Patton 1999; Valageas \& Silk 1999; Wu et al. 2000; Kravtsov \& Yepes 
2000). In particular, Valageas \& Silk propose high-redshift preheating of the 
IGM by radiation from quasars. It is quite natural that the mechanism which 
accounts for the required heating of the entire ICM affects also the evolution 
of its host.

In the context of the relationship between the QSO and the host galaxy, it is 
also worth noticing that the model predicts a metal abundance of the gas at 
the epoch of the QSO appearance in close agreement with the values inferred 
from the BELRs of high-$z$ QSOs.

In conclusion, the basic assumption of the model, namely that massive 
spheroidal galaxies evolve rapidly after the virialization of their haloes 
till their QSOs shine, is reinforced by the match of the model outcomes with 
the observed properties. In particular, in this paper we have shown that QSOs 
can play a major role in determing the observed chemo-spectrophotometric 
properties of spheroidal galaxies in a wide range of mass.

\section*{Acknowledgments}
L. S. and L. D. thank Gian Luigi Granato for the substantial contribution 
to the spectrophotometric code used for this work and for enlightening 
discussions. We also thank the referee for several comments and suggestions 
that improved the presentation of this work.

\label{lastpage}

\end{document}